\documentclass{aiaa-tc}

\usepackage{wrapfig}
\usepackage{threeparttable}
\usepackage{dcolumn}
\newcolumntype{d}{D{.}{.}{-1}}
\usepackage{nomencl}
\makeglossary
\usepackage{subfigure}
\usepackage{subfigmat}
\usepackage{fancyvrb}
\fvset{fontsize=\footnotesize,xleftmargin=2em}
\usepackage{lettrine}

\usepackage{booktabs}
\usepackage{bm, amssymb, amsmath, array, pdfpages}
\usepackage{amsmath,amssymb,color,graphicx,pdfsync,overpic,color,epstopdf,rotating,dashrule,float,bm}
\usepackage{float}

\usepackage{setspace}
\usepackage{comment}
\usepackage{changepage}

\usepackage{titlesec}

\graphicspath{{Figures/}}

\title{Data-Driven Low-Dimensional Modeling and Uncertainty Quantification for Airfoil Icing}

\author{%
  Anthony M. DeGennaro\thanks{Department of Mechanical and Aerospace
    Engineering; Student Member, AIAA.}
  \ ,
  Clarence W. Rowley\thanks{Department of Mechanical and Aerospace Engineering;
    Associate Fellow, AIAA.}
  \ , and
  Luigi Martinelli\thanksibid{2}\\
  {\normalsize\itshape
   Princeton University, Princeton, NJ, 08540, USA}\\
 }

 \AIAApapernumber{200?-????}
 \AIAAconference{Conference Name, Date, and Location}
 \AIAAcopyright{\AIAAcopyrightD{200?}}

\newcommand*{\mat}[1]{{\bf{#1}}}
\def\ip<#1,#2>{\left\langle #1,#2\right\rangle}

\begin{document}

\maketitle

\begin{abstract}
  The formation and accretion of ice on the leading edge of an airfoil
  can be detrimental to aerodynamic performance. Furthermore, the
  geometric shape of leading edge ice profiles can vary significantly
  depending on a wide range of physical parameters, which can
  translate into a wide variability in aerodynamic performance. The
  purpose of this work is to explore the variability in airfoil
  aerodynamic performance that results from variability in leading
  edge ice shape profile. First, we demonstrate how to identify a
  low-dimensional set of parameters that governs ice shape from a
  database of ice shapes using Proper Orthogonal Decomposition
  (POD). Then, we investigate the effects of uncertainty in the POD
  coefficients. This is done by building a global response surface
  surrogate using Polynomial Chaos Expansions (PCE). To construct this
  surrogate efficiently, we use adaptive sparse grid sampling of the
  POD parameter space. We then analyze the data from a statistical standpoint.

\end{abstract}



\section{Introduction}

Leading edge wing ice can present a significant safety hazard to
pilots. Wing ice shapes are often sharp and jagged in profile and
rough in surface texture; both of these characteristics tend to induce
flow separation at angles of attack which are often quite mild. This
can cause stall at lower angles of attack than those to which pilots
are accustomed. It is for this reason that wing icing has been
identified as the cause of 388 crashes between 1990 and 2000
\cite{landsberg}.

Equally troubling is the fact that wing icing is a process which is
associated with a high degree of uncertainty in practice. If one were
to analyze a database of ice shapes---regardless of whether the data
come from wind tunnels, flight tests, or computation---one would
usually find that the ice profile can vary significantly, depending on
a wide range of physical parameters (e.g., the liquid water content of
the freestream, airfoil geometry, the distribution of droplet
properties, etc.). It is therefore natural to ask how airfoil
aerodynamic performance is affected by variations in the ice
shape. This is the main question we hope to contribute towards
addressing in this work.

The approach we adopt is data driven, in the sense that all of the ice
shape variations that we will explore are, in a fundamental way,
derived from pre-existing databases of ice shapes. This is in contrast
to prior related studies, in which the ice shape variations that were
examined were generated in a somewhat heuristic fashion \cite{degennaro}.

The framework we adopt includes a combination of techniques from
low-dimensional modeling and uncertainty quantification. Specifically,
given a database of ice shapes, we first identify a low-dimensional
set of shape parameters by using the Proper Orthogonal Decomposition
(POD) of the dataset. This gives us both a method for generating and
studying artificial ice shapes as well as a way to assess the
statistical variation of our dataset. 

Based on the variation in the data, we select a parameterized range in
POD space to perform uncertainty quantification (UQ). We generate
artificial ice shapes by taking linear combinations of the POD modes,
and study the effects on aerodynamics by meshing the resultant airfoil
and computing its aerodynamic properties using an in-house flow solver. In order to perform the UQ
efficiently, we fit a surrogate function to the data using a
polynomial basis that is orthogonal to the input parameter
probability distribution; this method is known as Polynomial Chaos
(PC). We use an adaptive sparse grid algorithm as our collocation
method.

\section{Uncertainty Quantification for Airfoil Icing: Methodology}

In this section, we describe the various techniques we apply. Specifically, we first show
how we identify low-dimensional models for icing datasets. These
models will form the input parameter space for our UQ studies. Next,
we discuss the flow solver we will be using for the iced airfoil
calculations. The aerodynamic quantities we derive from these flow
calculations form the response function outputs in our UQ
studies. Finally, we describe the methods we will be using to quantify
uncertainty in our response function, as a result of uncertainty in
our input parameter space.

\subsection{Low-Dimensional Modeling using POD}

The space of all possible ice shapes contains an infinite number of degrees of
freedom.
However, for exploring parameter space, we wish to restrict our study
to a relatively small number of parameters that describe {\em likely\/} ice
shapes.
The approach taken here is to consider a subspace of possible ice shapes,
determined from data, either from simulations~(\S \ref{sec:results-sim}) or
experiments (\S \ref{sec:results-expt}).
In particular, for a given dataset of ice shapes, we determine a low-dimensional
subspace that optimally spans the data using Proper Orthogonal Decomposition
(POD)~\cite{holmes}.

Denote the vector of parameters governing a particular ice shape by $x \in \mathbb{R}^N$ (we
will have more to say about what $N$ is and how we get it soon).  We
wish to approximate any $x$ as a linear combination of some basis vectors
$\psi_i$, called POD modes:
\begin{equation}
x \approx \sum_{i=1}^P a_i \psi_i
\end{equation}
POD gives us one way of generating the basis $\{\psi_i\}$ from the data. A
key property is that the basis identified by POD will be able to
represent the dataset better than any other linear basis, in
the sense of projection error (using the standard Euclidean
norm).
The POD modes are determined as follows.  Let $x_j$, $j = 1 \ldots M$
denote the elements in our dataset, and let $\mat{X} \in \mathbb{R}^{N
  \times M}$ be the matrix with $x_j$ as its columns:
\begin{equation}
\mathbf{X} =
 \begin{bmatrix}
   \vline & & \vline \\
   x_1 & \cdots & x_M \\
   \vline & & \vline \\
 \end{bmatrix} 
\end{equation}
The POD modes $\psi_i$ are then given by the left singular vectors of
the singular value decomposition $\mat{X} =
\mat{U}\mat{\Sigma}\mat{V}^T$ (i.e., the columns of $\mat{U}$).

\subsection{Flow Solver Description}

In order to evaluate the aerodynamic characteristics for the different ice
shapes considered, we
need a reliable, tested flow solver. In this paper, we use use FLO103,
a well-validated, in-house code for the solution of the
two-dimensional Reynolds Averaged Navier-Stokes (RANS) equations
developed over the course of many years by Martinelli and
Jameson~\cite{Martinelli:Validation,Tatsumi:High,martinelli}. This code has
also been used to perform similar aerodynamic calculations for iced
airfoils in a previous study~\cite{degennaro}.

A one-equation Spalart-Allmaras turbulence model~\cite{Spalart:One}
provides closure for the RANS, which is capable of accurately modeling
mildly separated flow near the stall regime.  The discretization of
the spatial operators is carried out by using a second order
cell-centered finite-volume method in which the viscous stresses are
computed using a discrete form of Gauss' theorem.  The key to the flow
solver efficiency is a full approximation multigrid time-stepping
scheme, which accelerates the rate of convergence to a steady state.

\subsection{Uncertainty Quantification using PCE}

The POD modeling of the ice dataset has---viewed from a UQ
perspective---generated for us a relatively low-dimensional parameter space to explore. Casting
this as a UQ problem has several potential benefits and uses:
\begin{itemize}
\item A linearly parameterized description of the ice gives us an easy
  and systematic method of generating new ice shapes that effectively
  interpolate the shapes present in our database. This makes it
  possible to produce and study a wide range of shapes.
\item UQ tools allow us to compute and analyze the statistical
  relationships between our responses (aerodynamics) and our inputs
  (ice shapes). For example, we can look at the effect of different
  input distributions (eg. uniform, Gaussian, etc.), correlations
  between POD modes and lift coefficients, output statistics, etc.
\item The UQ tools we will be using produce for us a surrogate model
  of the input-output behavior -- we obtain a polynomial mapping
  between the POD modes and the aerodynamics. This can be advantageous
  as a predictive tool: if one wishes to know the aerodynamics of a
  particular ice shape that has not been studied yet, one could
  compute it by simply evaluating the surrogate model (assuming this
  ice shape is in the span of the POD basis).
\end{itemize}

Furthermore, using POD to generate our parameterization of the ice is
advantageous for a few reasons:
\begin{itemize}
\item The POD basis outperforms any other possible linear basis of the
  same dimension for representing our data in the sense of projection
  error. In this way, it is an optimal parameterization.
\item The POD coordinates are linearly uncorrelated (since the modes are
  orthonormal). This justifies an assumption of mutual independence
  amongst the parameters in our UQ study, which underlies the UQ
  methods we will be using.
\item The POD is a general method for data-processing, which makes it
  amenable to analyzing other ice shape databases that we have not
  yet considered. Thus, our approach could be applied to other
  datasets as well.
\end{itemize}

Here, we detail exactly how our ice shape modeling can be
formulated as a parameterized UQ study. We then present results for
such a study, and analyze the statistical information that we can
glean from it.

As previously noted, the UQ methods that we will be using need to be
efficient, since we have a moderately high dimensional parameter space
to explore. Additionally, we would like to obtain a surrogate model
that explicitly describes the dependence of the airfoil aerodynamics
on ice shape, which can be inexpensively sampled and analyzed for
statistical correlations. As a result of these requirements, we choose
to use Polynomial Chaos Expansions (PCE) as our framework for
performing UQ.

We give a brief overview of this approach below; further details can
be found in introductory references on PCE
methods\cite{ghanem_book,xiu_book,lemaitre}.

Let~$Z=(Z_1,\ldots,Z_d)$ be a vector of random variables that
parameterize the uncertain quantities in the ice shape. We are
interested in the corresponding uncertainty of an aerodynamic
quantity, represented by $y(Z)$.  In our setting, $Z$ are the POD
coefficients, and $y(Z)$ will be the airfoil lift and drag
coefficients at a particular angle of attack, computed by FLO103.

The goal of the method is to represent $y(Z)$ in terms of some basis
functions~$\Phi_i$. Assuming (for ease of exposition) that $y(Z)$ is
scalar-valued, we write:
\begin{equation}
  \label{eq:1}
  y(Z) = \sum_{|i|=0}^N y_i \Phi_i(Z).
\end{equation}
Here, $i=(i_1,\ldots,i_d)$ is a multi-index, and $|i|=\sum_{j=1}^d i_j$.  We define an inner
product on the space of functions of the random variables by
\begin{equation}
  \label{eq:2}
  \ip<f,g> = \int_\Gamma f(Z) g(Z) \rho(Z)\,dZ,
\end{equation}
where $\rho(Z)$ denotes the probability density function of $Z$, and
has support $\Gamma$.  A fundamental insight in PCE methods is to
employ basis functions that are orthonormal with respect to this
inner product, so that
\begin{equation}
  \label{eq:3}
  \ip<\Phi_i,\Phi_j> = \delta_{ij},
\end{equation}
where $\delta_{ij}=1$ if $i=j$, and $0$ if $i\ne j$. In particular,
a multivariate basis polynomial $\Phi_i$ may be written as
\begin{equation}
\Phi_i(Z) = \prod_{k=1}^d \phi_{i_k}(Z_k),
\end{equation}
where $\phi_n$ is a (univariate) polynomial of degree~$n$. The $\{ \phi_n\}$ will be a basis of
orthogonal polynomials chosen so that the orthogonality
condition~\eqref{eq:3} is satisfied. In this paper, we work
exclusively with uniformly distributed random variables, and so our
basis polynomials are the multivariate Legendre polynomials.

The coefficients $y_i$ in the expansion~\eqref{eq:1} may be determined
by taking an inner product with $\Phi_j$: because the $\Phi_j$ are
orthonormal, we have
\begin{equation}
  \label{eq:4}
  y_j = \ip<y,\Phi_j>.
\end{equation}
Note that one could also take $y(Z)$ to be a vector of several different
aerodynamic quantities of interest: in this case, the coefficients~$y_i$ in the
expansion~\eqref{eq:1} are vectors, and each component of $y_i$ is determined by
an equation such as~\eqref{eq:4}, for the corresponding component of~$y$.

The important issue now is how we choose to approximate the projection
integrals in~\eqref{eq:4}. A possible choice is to use Gauss
quadrature, in which the function $y(Z)$ is evaluated on a grid
consisting of the tensor product of $n$ separate 1-D quadrature point
sets in parameter space. However, this method suffers from the
curse of dimensionality, since the number of required samples grows
exponentially with the dimension~$n$.

A commonly used alternative is to use sparse grid methods\cite{Smolyak}, in
which the number of grid points used is lessened by using only a subset of the
full tensor product.
Another advantage is that anisotropic
adaptive $p$-refinement of the mesh is possible, since nested 1-D
nodes are used. In an adaptive setting, global sensitivities are
calculated using total Sobol indices, which are defined for each
parameter as:
\begin{equation}
T_i = \frac{\mathbb{E} [ Var(y|Z_{-i}) ]}{Var(y)},\qquad i = 1,\ldots,d,
\label{eq:sobol}
\end{equation}
Here,
$Var(y|Z_{-i})$ denotes the variance of $y(Z)$ given all parameters
except $Z_i$. This is a measure of ``how much'' parameter $Z_i$
contributes to the total variance of $y(Z)$ on average. Parameters
that have higher Sobol indices contribute more to the variance of the
response and hence require more refinement than parameters with lower
Sobol indices.

Further details on the sparse grid approach can be found in standard
references\cite{lemaitre,Gerstner:SparseGrids}. In our study, we compute the
sparse grids using DAKOTA, an
open-source code for optimization and UQ developed by Sandia National
Laboratory~\cite{Dakota}.

\section{Results: Ice Shapes from Simulations}
\label{sec:results-sim}

\begin{figure}[htb]
\centering
\includegraphics[width=0.8\textwidth]{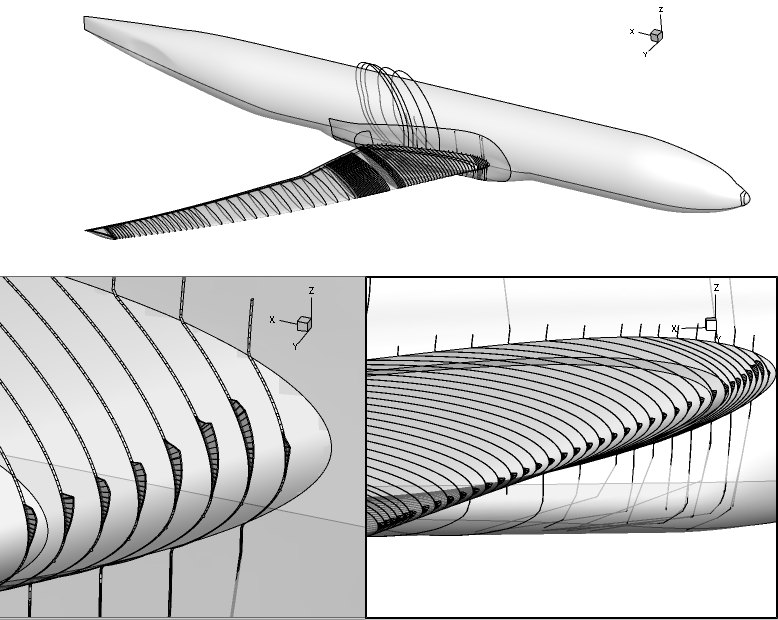}
\caption{Horn ice accretion case on the NASA CRM, from Broeren et. al.\cite{broeren}.}
\label{fig:CRMHorn}
\end{figure}

In this section, we give our first example of how the techniques just
discussed may be applied to an ice shape dataset from the
literature. For this example, our data consists of cross-sections
from an icing simulation performed on a 3D swept wing. This data and
the research related to it is described in detail in Broeren
et. al. \cite{broeren} and is shown in Figure \ref{fig:CRMHorn}.

The model geometry used in this data set was the NASA Common Research
Model (CRM) at 65\% scale. The ice accretion model used was LEWICE3D,
which is a NASA's 3D icing code.  This study represents 45 minutes of
accretion time at an altitude of 10,000 ft and free stream velocity of
232 knots. The static temperature was $-4^\text{o}$ C, the mean
volumetric diameter (MVD) was 20 $\mu$m, and the liquid water content
(LWC) was 0.55 g/m$^3$.

\subsection{Modeling using POD}

Our first objective is to apply the POD machinery to the data set
consisting of 95 individual horn ice cross sections on the CRM wing,
with the objective of identifying the most important ice shape
features. We will first do a preprocessing step on our data -- we will
map all of the airfoil cross sections into $s$-coordinates (arc length
coordinates, relative to the airfoil leading edge), so that the ice
shapes may be represented as a function of a single variable.

Let us denote the horn ice height (in $s$-coordinates, normal to the
airfoil) as $N_k(s)$, where $k$ indexes the cross-sections.  A POD
representation of this dataset will take the following expansion
form:

\begin{equation}
N_k(s) = \overline{N}(s) + \sum_{i=1}^M c_{i_k} \psi_i(s),\qquad
k=1,\ldots,S.
\end{equation}

Here, $\overline{N}(s)$ is the height (in s-coordinates) of the mean
ice shape, $c_i$ is the $i^{th}$ POD coefficient, and $\psi_i(s)$ is
the $i^{th}$ POD mode.

\begin{figure}[htb]
\centering
\subfigure[Horn profiles unaligned.]{\includegraphics[width=.4\textwidth]{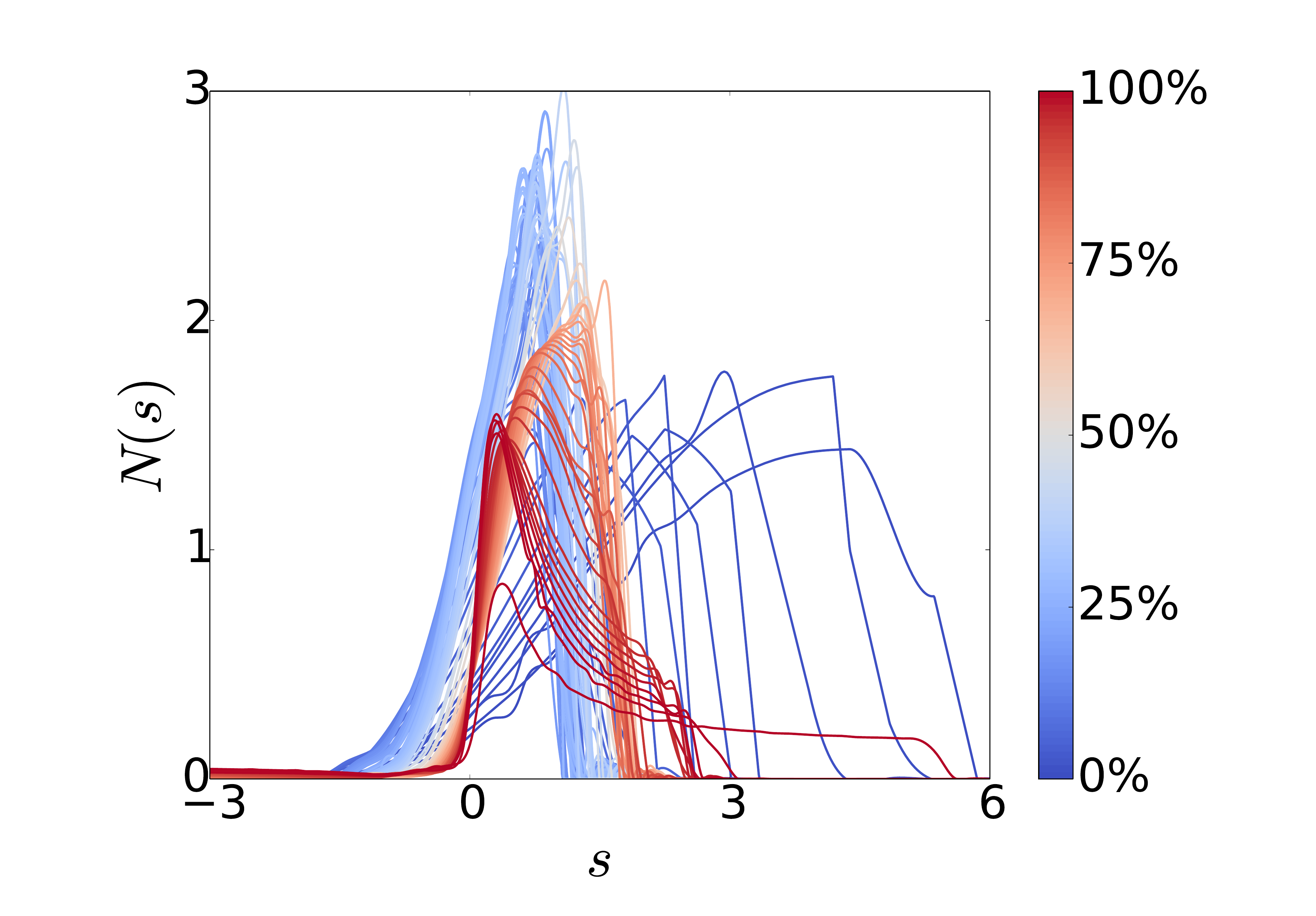}}
\subfigure[Horn profiles scaled/shifted to fit a symmetric Gaussian template centered at zero.]{\includegraphics[width=.4\textwidth]{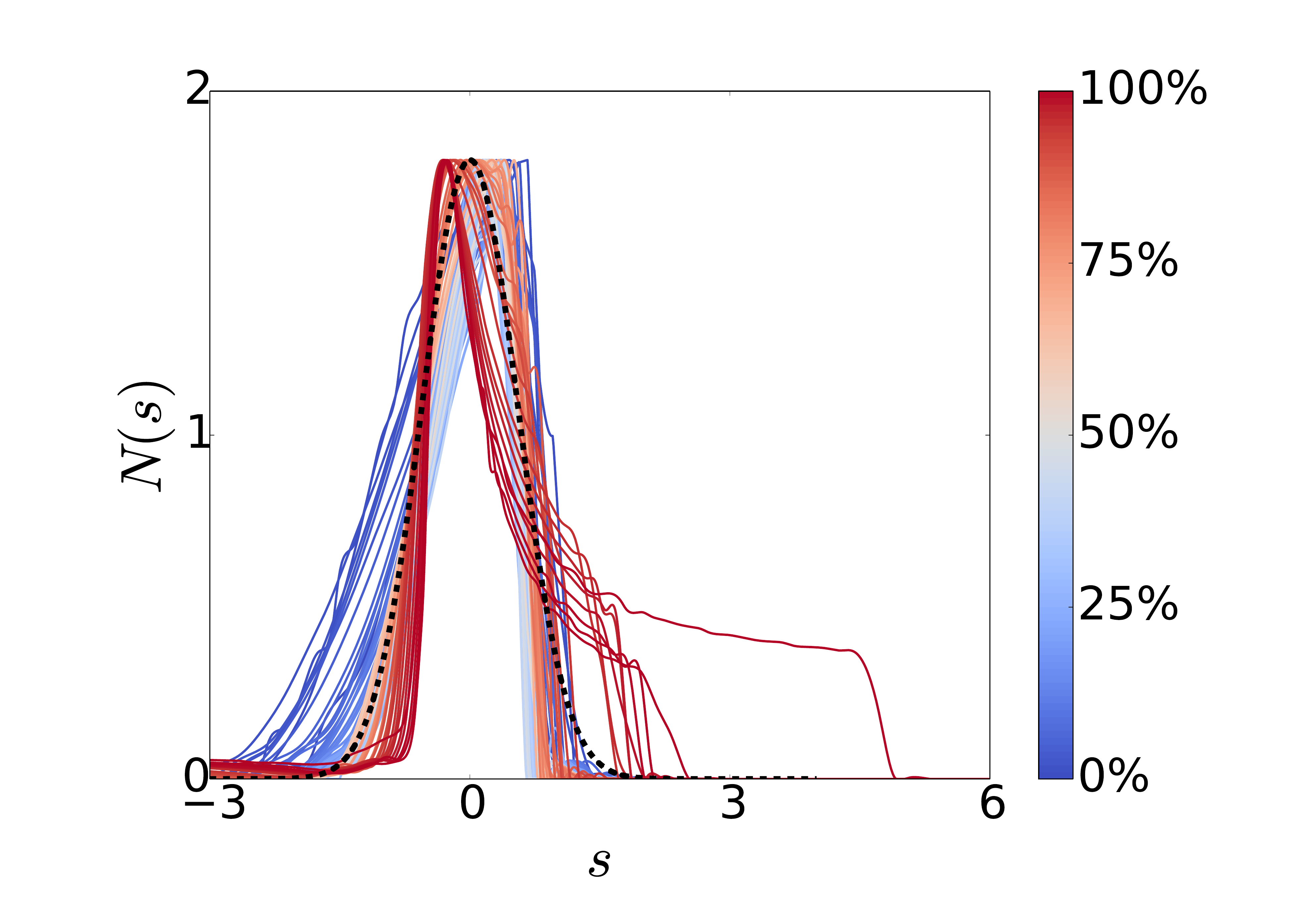}}
\caption{Horn profiles, unaligned and aligned. The color map denotes spanwise position along the CRM wing from root (0\%) to tip (100\%).}
\label{fig:HornsAligned}
\end{figure}

Fig.~\ref{fig:HornsAligned}(a) shows the dataset. By inspection, we
see that much of the variability in shape is due to differences in
width, position, and height. We can account for this by
scaling/shifting each of the individual ice shapes to fit a template
shape, which produces Fig.~\ref{fig:HornsAligned}(b). The point of
doing this is to separate variations in {\it scaling} from differences
in {\it shape}. As we will see, this will be reflected in our
parameterization of the ice, in which three of the parameters specify
scaling, and the two POD modes specify shape perturbations.

If we pre-process the data with these scalings/shiftings, the POD expansion will be:

\begin{equation}
\label{eq:5ParamPOD}
N_k(s) = h_k \left( \overline{N}(a_k s + b_k) + \sum_{i=1}^M c_{i_k} \psi_i(a_ks + b_k) \right),\qquad
k=1,\ldots,S.
\end{equation}

In this framework, uncertainty could be accounted for by perturbing the POD coefficients:

\begin{equation}
\begin{aligned}
a_k & \mapsto a_k + \xi_a \\
b_k & \mapsto b_k + \xi_b \\
h_k & \mapsto \xi_h h_k \\
c_{i_k} & \mapsto c_{i_k} + \xi_{c_i} \;\;\; \text{for $k$ = 1 ... $M$}\\
\end{aligned}
\end{equation}

Here, the first three parameters are perturbations on the nominal
positions, widths, and heights of all of the individual ice
shapes. The last $M$ parameters are global perturbations on each of
the POD eigenmodes. The first three parameters were chosen by
scaling/shifting each individual shape so that the transformed shape most
closely matches a symmetric Gaussian
template~$G(s)$ (which is close to the mean of the unshifted/unscaled
data):
\begin{equation*}
(a_k,b_k) = \arg\min_{a,b} \| N_k(as+b) - G(s) \|_2^2,\qquad k = 1,\ldots,S.
\end{equation*}

\begin{figure}[htb]
\centering
\subfigure[Magnitude of POD eigenvalues.]{\includegraphics[width=0.4\textwidth]{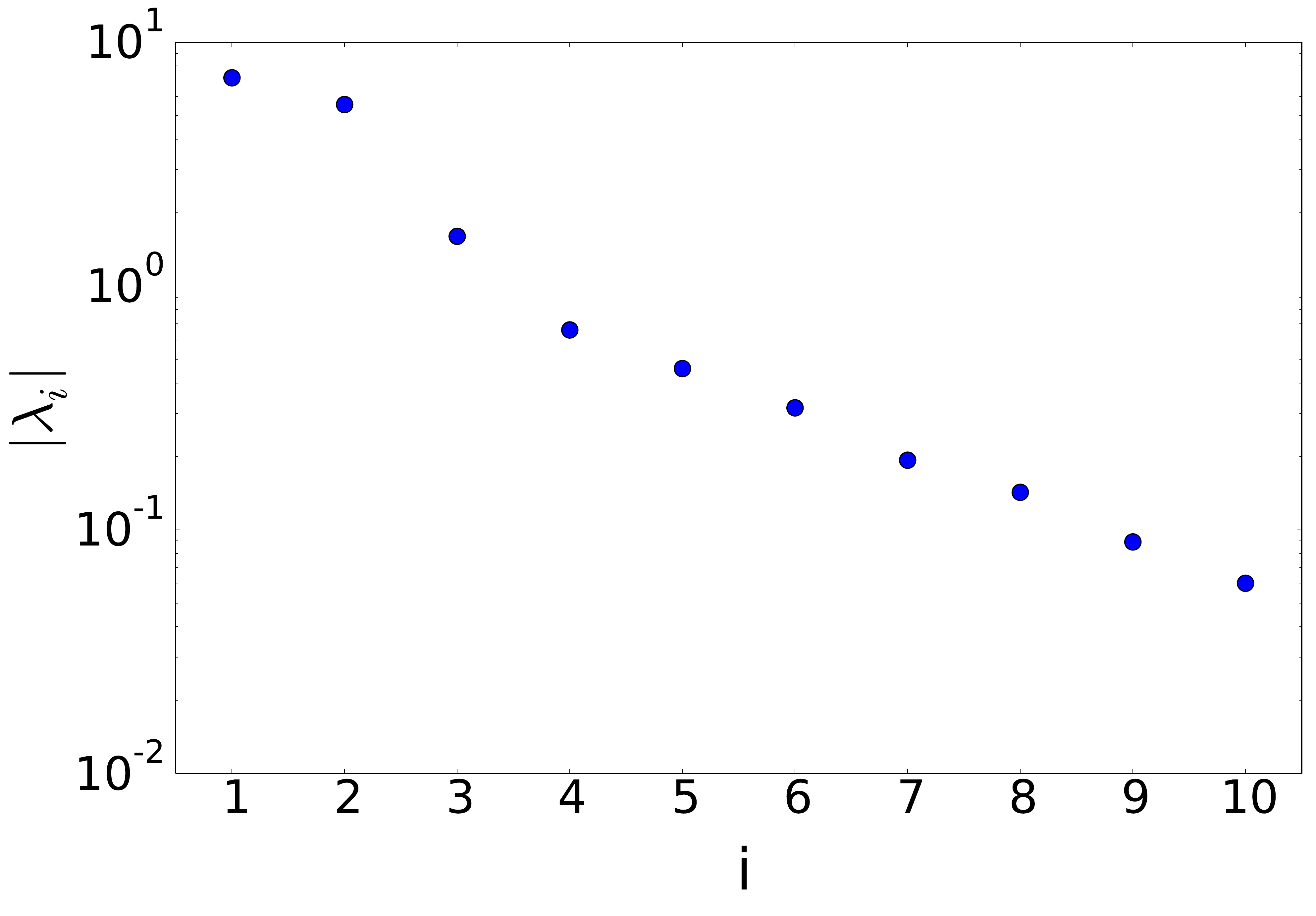}}
\subfigure[The mean and first 2 POD modes.]{\includegraphics[width=.4\textwidth]{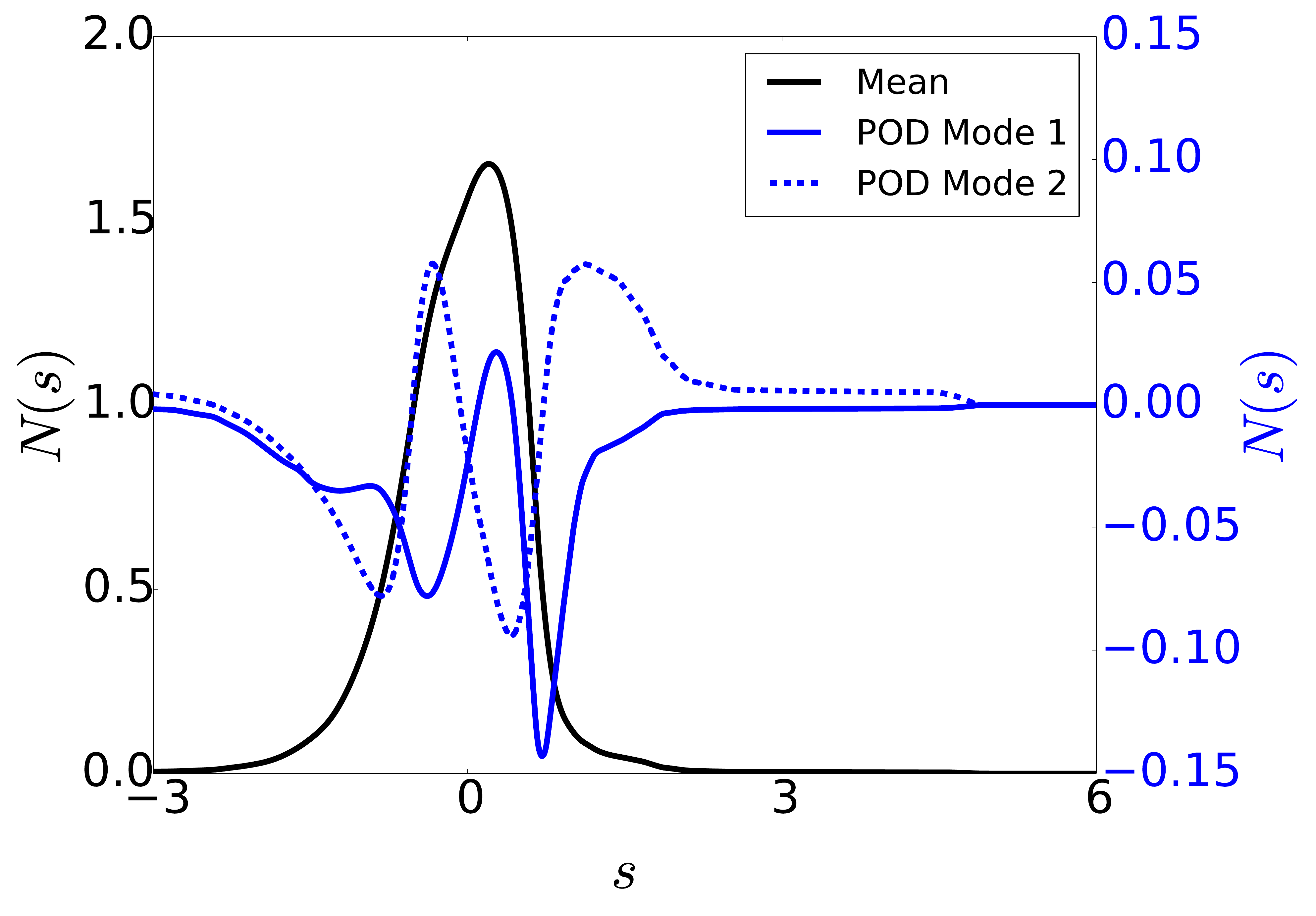}}
\caption{POD eigenvalues and modes for the scaled/shifted data.}
\label{fig:PODModesEvals}
\end{figure}

The absolute values of the POD eigenvalues are shown in Figure
\ref{fig:PODModesEvals}.  As can be seen, there is a sharp drop-off in
the magnitude of the scaled/shifted POD eigenvalues at mode
2. This implies that the first two modes capture much of the important
features.

\begin{figure}[h!tb]
\centering
\subfigure[Reconstruction (1 POD mode).]{\includegraphics[width=.4\textwidth]{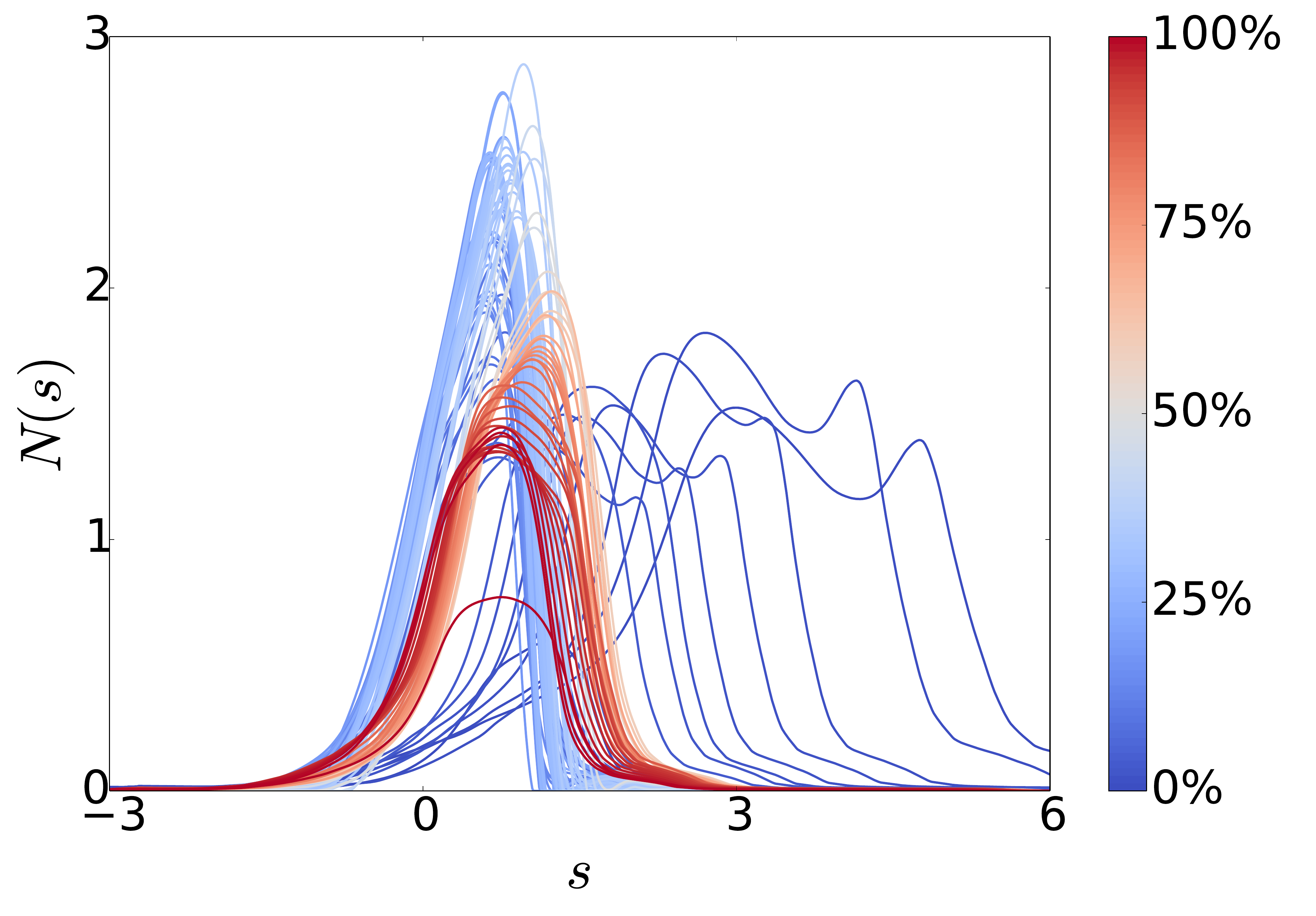}}
\subfigure[Reconstruction (2 POD modes).]{\includegraphics[width=.4\textwidth]{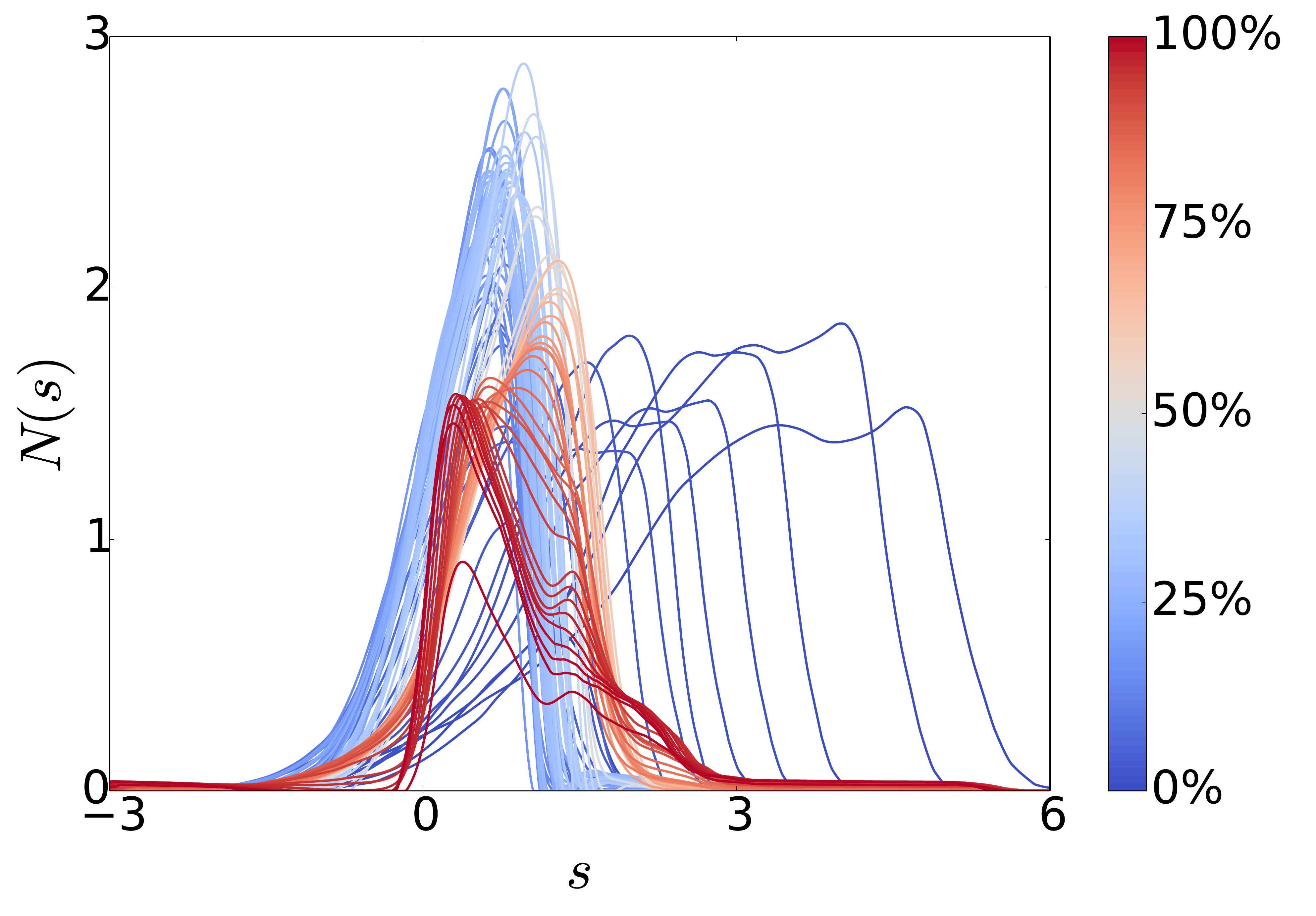}}
\caption{POD reconstructions of the ice. The color map denotes spanwise position along the CRM wing from root (0\%) to tip (100\%).}
\label{fig:ActualandReconstructions}
\end{figure}

Shown in Figure \ref{fig:ActualandReconstructions} are original ice
shapes along with their reconstructions using 1 and 2 POD modes. It
can be observed that no skewness is able to be represented until 2 POD
modes are used.

\begin{figure}[htb]
\centering
\includegraphics[width=0.5\textwidth]{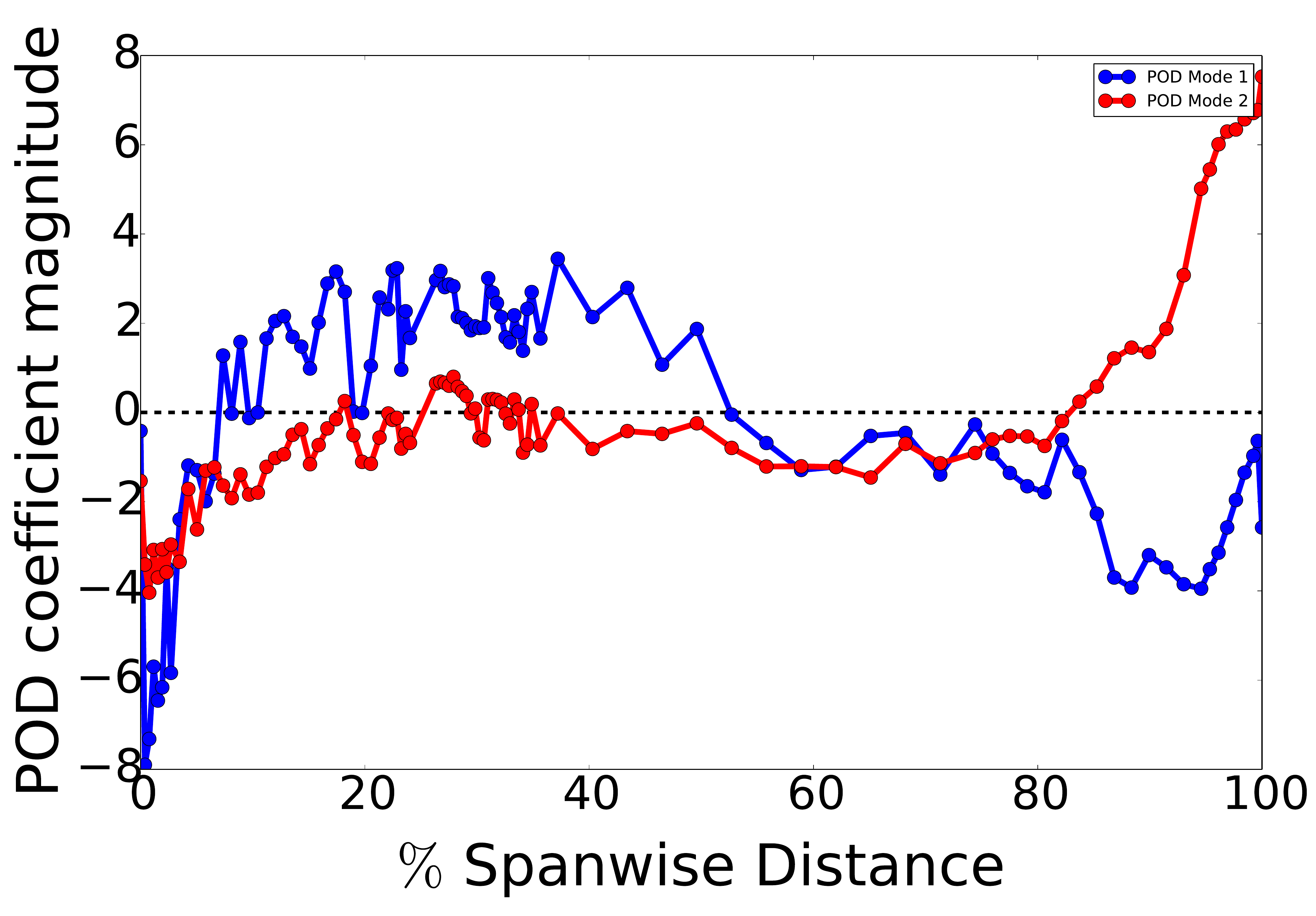}
\caption{Spanwise variation of the POD modes.}
\label{fig:SpanwiseModeVariation}
\end{figure}

\begin{figure}[h!tb]
\centering
\subfigure[Variation of the $1^{st}$ POD mode $\in (-2, 2)$.]{\includegraphics[width=.4\textwidth]{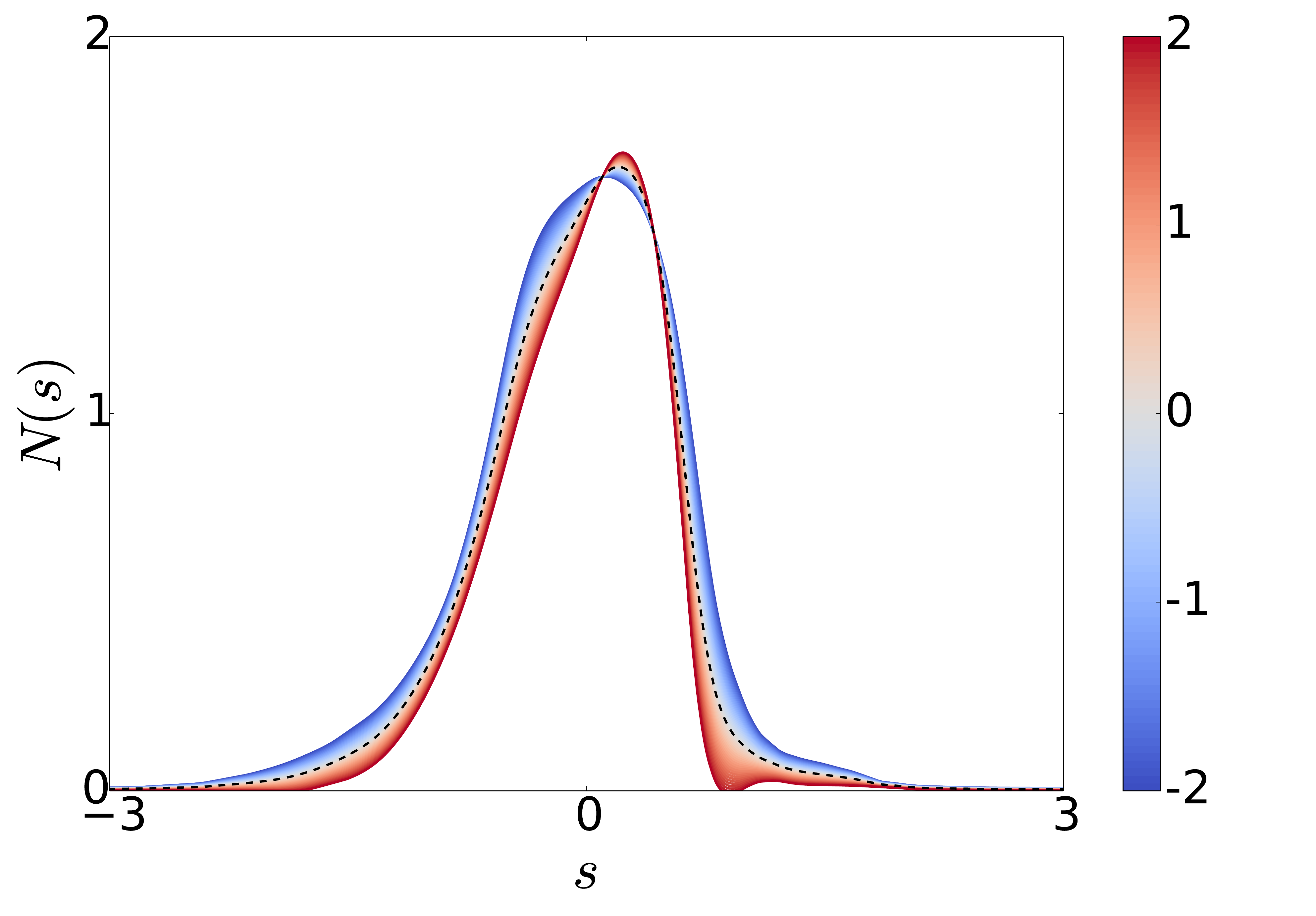}}
\subfigure[Variation of the $2^{nd}$ POD mode $\in (-2, 2)$.]{\includegraphics[width=.4\textwidth]{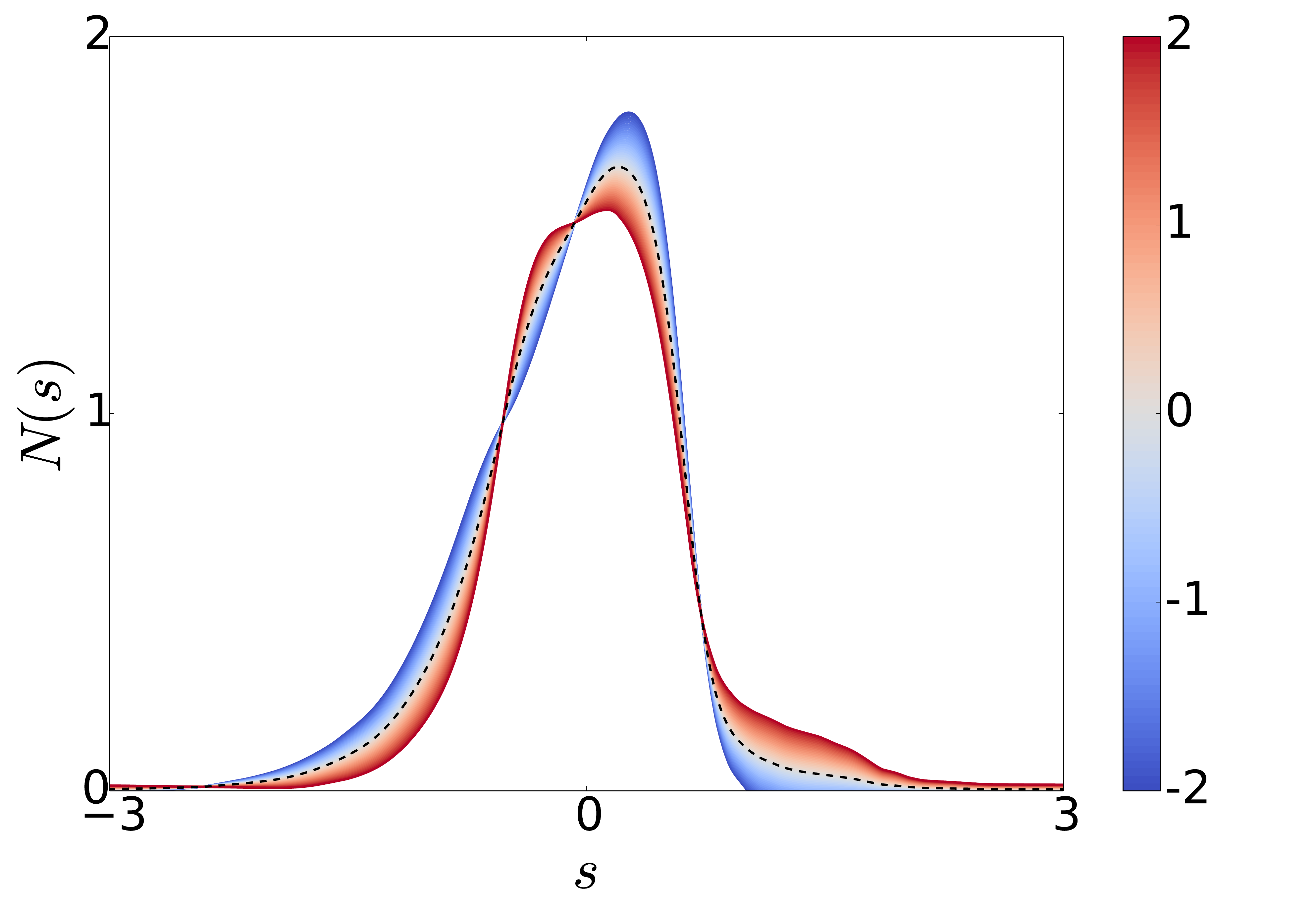}}
\caption{Variation of the POD modes on the mean (dashed black).}
\label{fig:VaryPODModes}
\end{figure}

The next step in identifying the relevant parameters for a UQ study is to study the 
spanwise variation of the POD coefficients. Figures \ref{fig:SpanwiseModeVariation} 
and \ref{fig:VaryPODModes} demonstrate these properties. As can be seen, 
both modes are most pronounced at the boundaries, since this is where ice shapes 
deviate the most from the mean. The first mode has the effect of making the ice 
profile wider and skewed left; hence, it is largest in magnitude on the inboard 
portions of the wing. The second mode has the effect of skewing the ice profile right; 
hence, it is largest in magnitude on the outboard portions of the wing.

\subsection{Airfoil Icing UQ: Five-Parameter Scenario}

\begin{figure}[h!]
\centering
\subfigure[Variation of the width.]{\includegraphics[width=.4\textwidth]{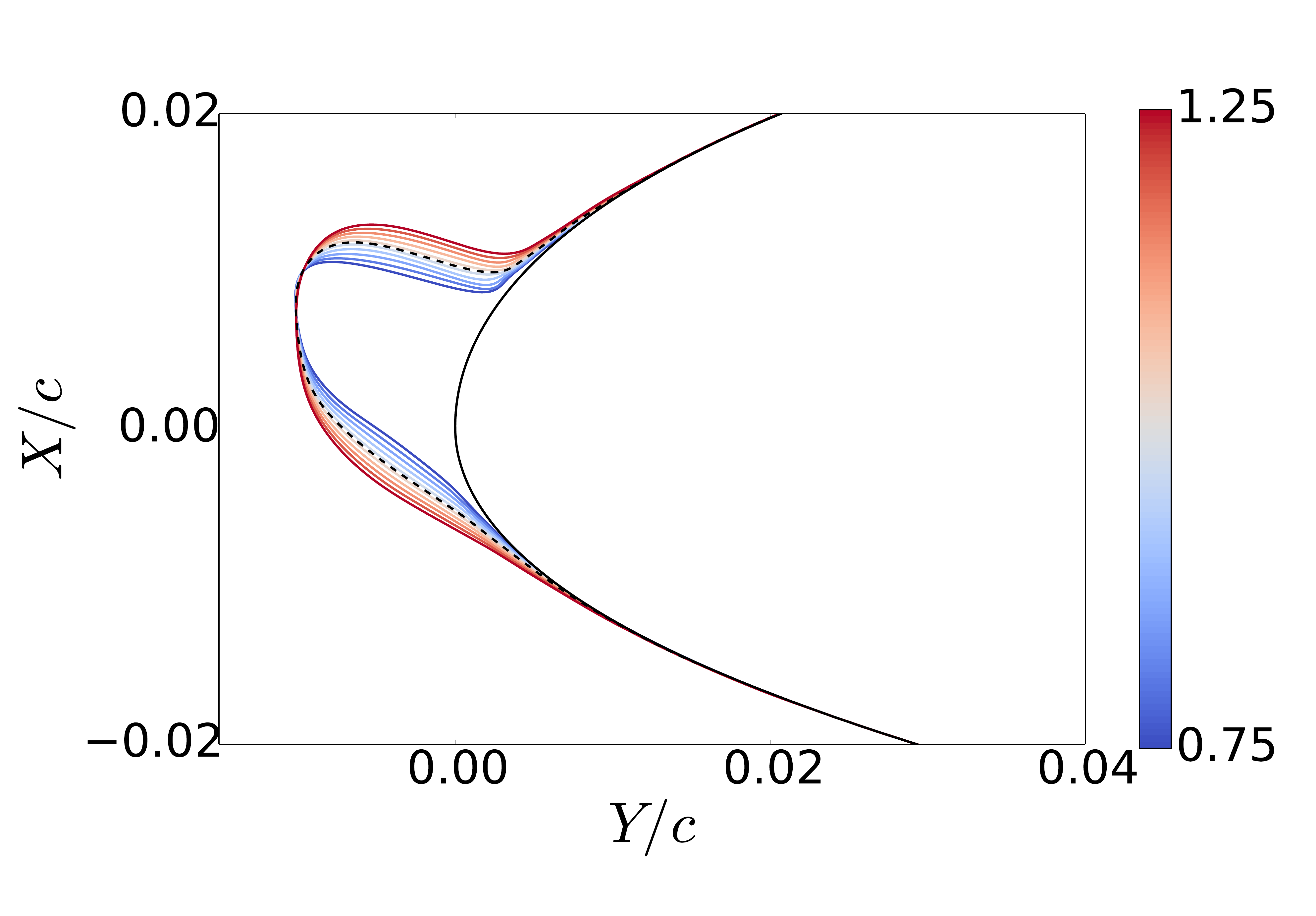}}
\subfigure[Variation of the height.]{\includegraphics[width=.4\textwidth]{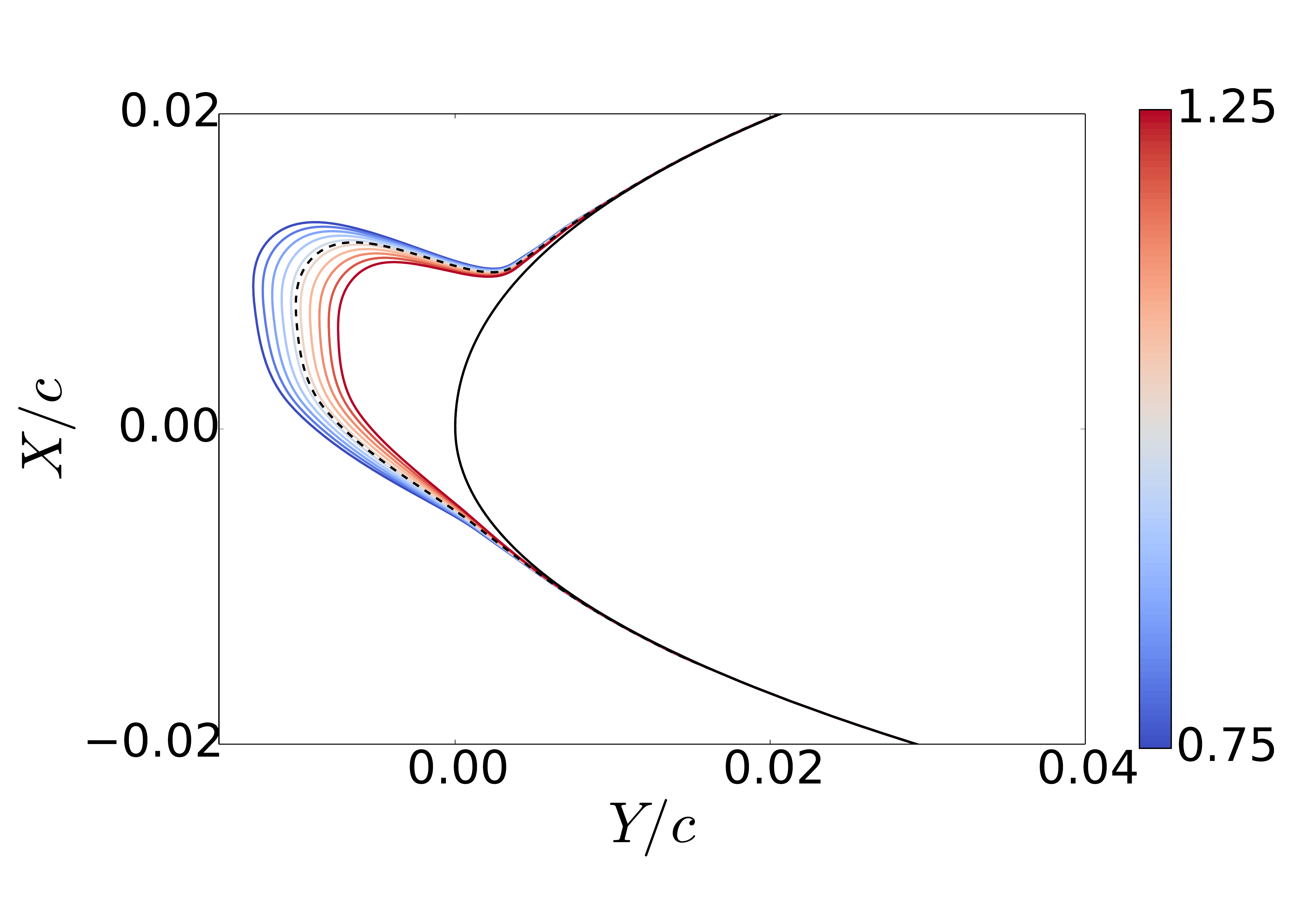}}
\subfigure[Variation of the position.]{\includegraphics[width=.4\textwidth]{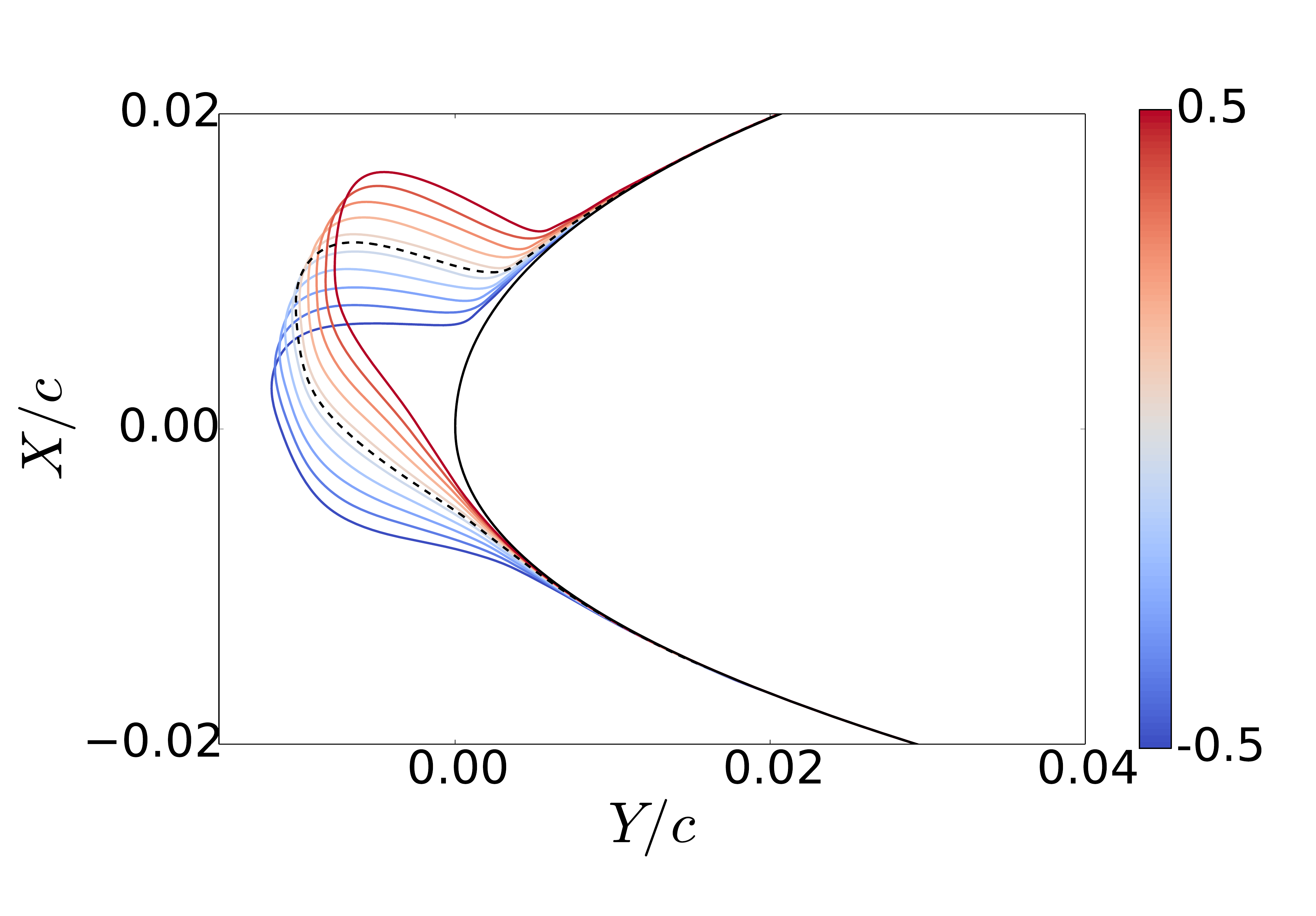}}
\subfigure[Variation of the POD modes. $Inset$: POD coordinates of the ice shapes (colored) and original data (black).]{\includegraphics[width=.37\textwidth]{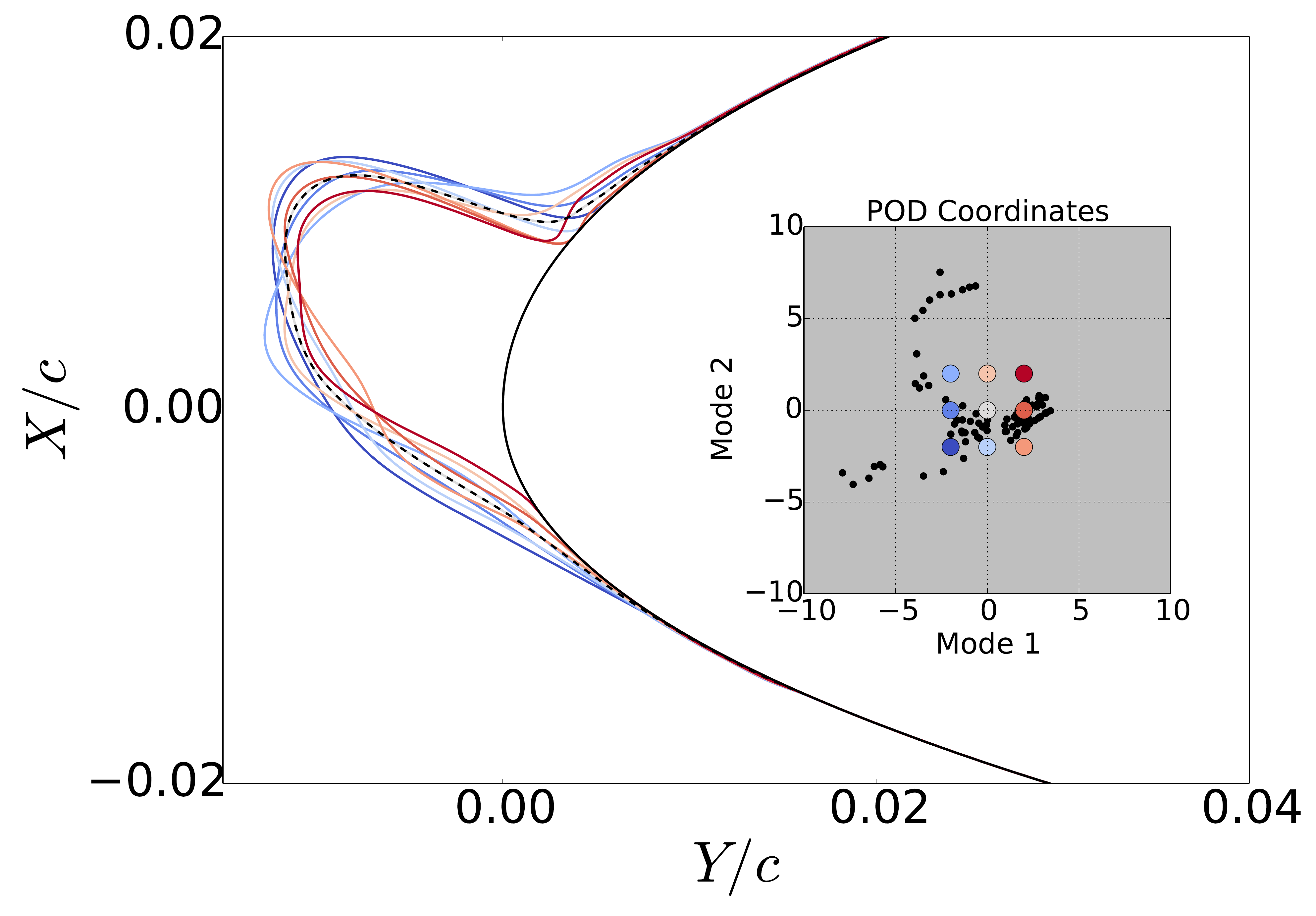}}
\caption{Depiction of the ranges of each of the parameters in the
  5-dimensional parameter space and their effects on the horn shape
  (colorbars indicate parameter ranges). Dashed black ice shape in
  each subplot represents the mean shape used.}
\label{fig:DifferentShapesPODModes}
\end{figure}

Now that we have generated a POD representation of our horn shape
data, we can investigate the effects of uncertainty in the POD
coefficients. As shown in Eq.~\eqref{eq:5ParamPOD}, we have 3 scaling
parameters (height, width, and position), and 2 POD coefficients, so
our parameter space is 5 dimensional. We assume that all 5 of our
parameters are uniformly distributed between some bounds. The ranges
that we choose, as well as their independent effects on the horn
shape, are displayed in Fig.~\ref{fig:DifferentShapesPODModes}.

\subsection{Results and Discussion}

Here we describe the results of the UQ investigation described in the
last section. We will quantify uncertainty in the two response metrics
$C_L$ and $C_D$. The airfoil cross-section used here was chosen to be
the cross-section at 50$\%$ semispan of the CRM, and the aerodynamic
coefficients were determined at a Reynolds number of $5 \times 10^6$, Mach
number of $0.4$, and
angle of attack $\alpha = 3^{\circ}$.

\begin{figure}[h!]
\centering
\subfigure[PDF for $C_L$.]{\includegraphics[width=.45\textwidth]{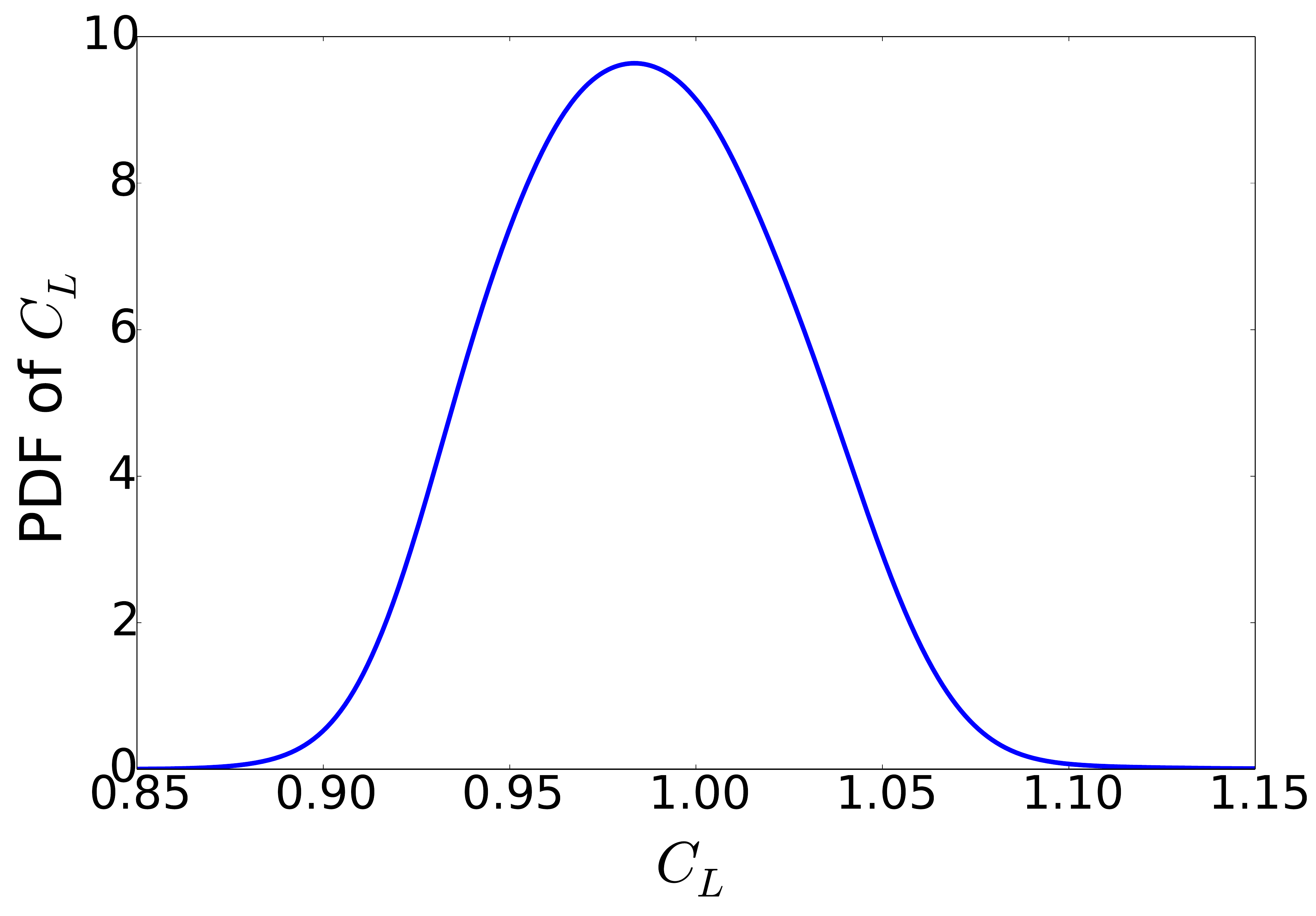}}
\subfigure[PDF for $C_D$.]{\includegraphics[width=.45\textwidth]{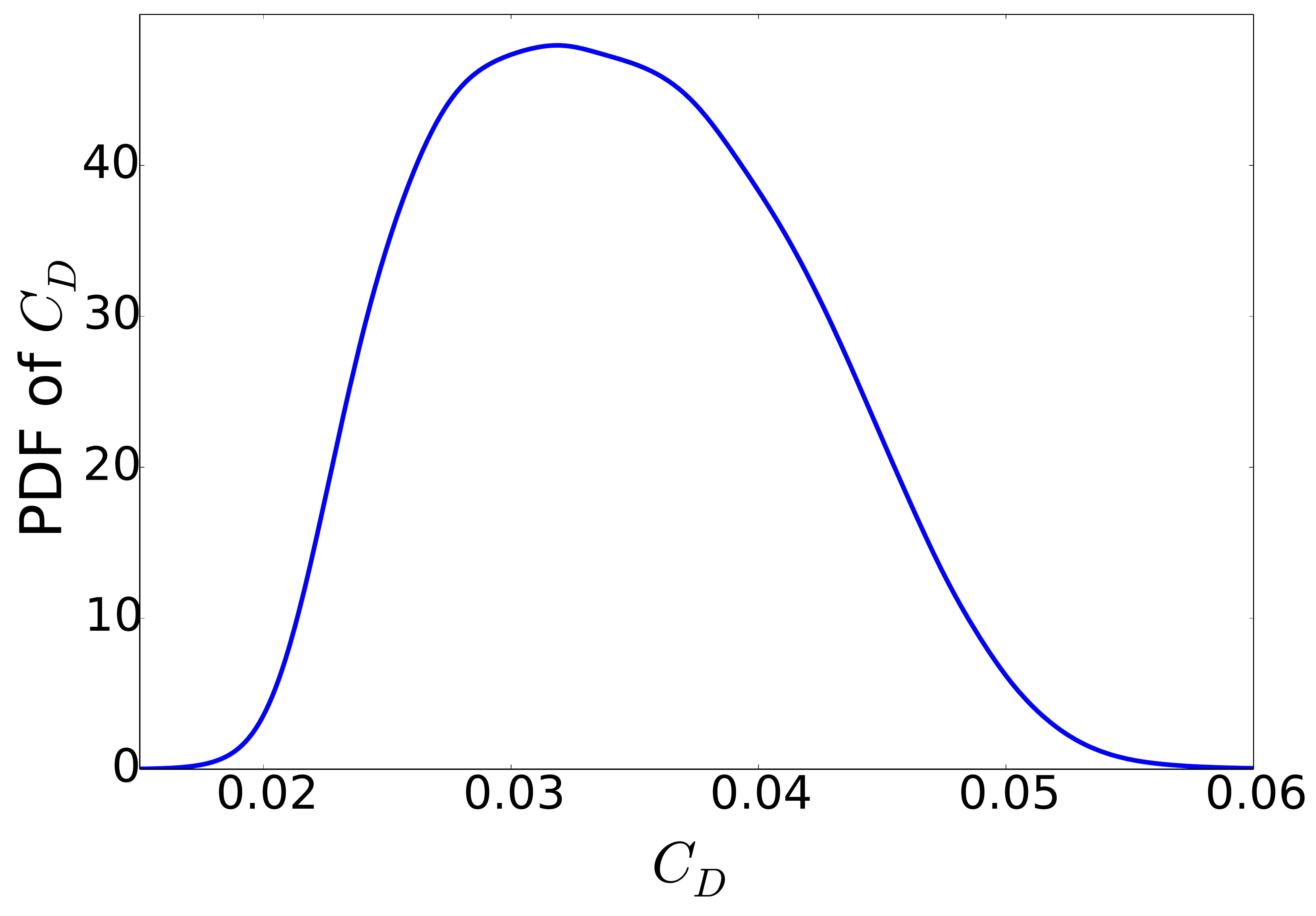}}
\caption{Probability density functions for two response metrics in the
  5 parameter UQ study. These results required 1,103 samples in the
  adaptive sparse grid algorithm.}
\label{fig:PDFResults}
\end{figure}

\newcommand{\ra}[1]{\renewcommand{\arraystretch}{#1}}
\newcommand{\squeezeup}{\vspace{-.5in}}
\begin{table*}[h!]\centering
  \caption{Data correlations}
\ra{1}
\begin{tabular}{crr}\toprule
& \hfill$C_L$\hfill & \hfill$C_D$\hfill\\ \midrule
$C_L$   & $1.00 $   & $-0.94$   \\
$C_D$   & $-0.94$   & $ 1.00$   \\ \midrule
$a$     & $0.09 $   & $-0.05$ \\
$b$     & $-0.78$   & $0.82 $  \\
$h$     & $-0.28$   & $0.31 $  \\
POD 1   & $-0.28$   & $ 0.26$   \\
POD 2   & $0.33 $   & $-0.34$   \\
\bottomrule
\end{tabular}
\label{tab:corr}
\end{table*}

\begin{table*}[h!]\centering
  \caption{Sobol Indices (Single Parameter)}
\ra{1}
\begin{tabular}{r c c c c c}\toprule
    & $a$    & $b$   & $h$   & POD 1 & POD 2 \\ \midrule
$T$ & 0.03   & 0.69  & 0.15  & 0.11  & 0.14 \\
\bottomrule
\end{tabular}
\label{tab:sobol}
\end{table*}


Fig.~\ref{fig:PDFResults} presents the statistics for a Latin
Hypercube sampling of the PCE surrogate with $10^6$ samples. This
surrogate required 1,103 flow solver evaluations to converge, which is
reasonable for a 5 dimensional parameter space. Convergence is
based on the change in the $L_2$ norm of the surrogate response
covariance matrix falling below some threshold (which we set to be
equal to $10^{-4}$).

The first thing to note is that we can easily see that $C_L$ and $C_D$
correlate very strongly (corr$(C_L,C_D) = -0.94$), which indicates
that over our parameter range, we only need to examine one of the two
in order to understand the effects of our inputs on aerodynamic
performance.

In order to examine the independent relative contributions of each of
our 5 input parameters to the variance of our responses, we examine
both the data correlations and the Sobol indices (defined previously
in Eq.~\eqref{eq:sobol}). Loosely speaking, the Sobol index gives a
measure of how much, on average, a parameter (or a combination of
parameters) contributes to the total variance of the response. It is
clear from these metrics that variations in horn position, $b$,
contribute the most to the variance of our response and hence $b$ is
the ``most important'' parameter. The caveat of this statement, of
course, is that it only applies over the limited parameter range we
have chosen. Had we chosen to investigate larger variations in height,
for example, then height could very well be the dominant parameter.

\begin{figure}[htb]
\centering
\subfigure[Lower surface horn.]{\includegraphics[width=.45\textwidth]{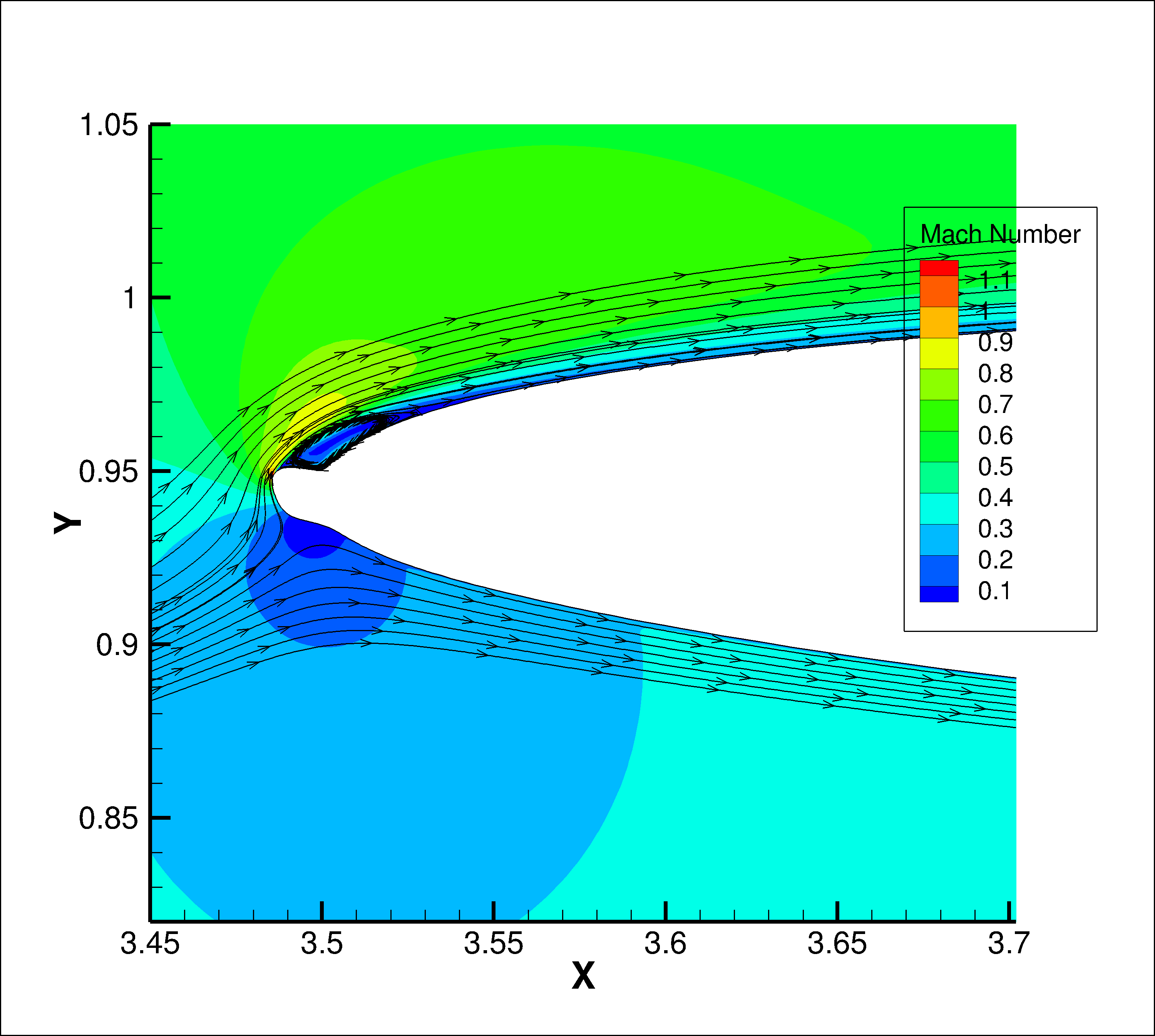}}
\subfigure[Upper surface horn.]{\includegraphics[width=.45\textwidth]{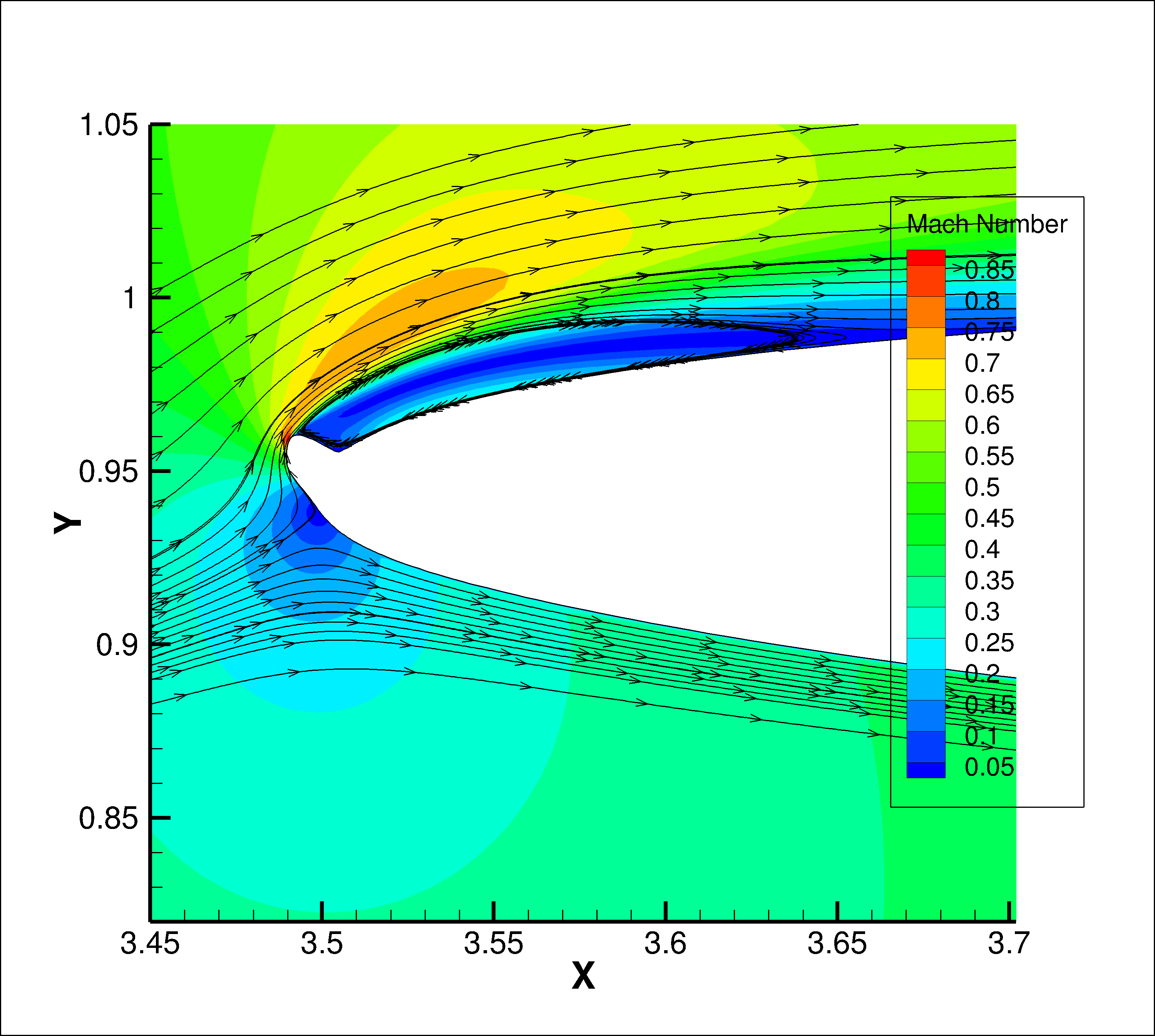}}
\caption{Flow field contours (of Mach number) for two horns of equal
  size and shape that differ only in their relative positions. The
  horn in ($b$) is more normal to the freestream flow than the horn in
  ($a$), and hence generates a larger scale leading edge separation
  bubble.}
\label{fig:GoodBadHorns}
\end{figure}

The sign of the correlation of horn position with our responses
indicates that performance degrades (ie., lower lift, higher drag) as
the horn moves closer to the upper surface. The physical explanation
for this is intuitive, and can ascertained by inspecting
Fig.~\ref{fig:GoodBadHorns}. The upper surface horn is a more
obtrusive flow obstacle, and therefore promotes a larger steady-state
separation bubble aft of the horn than the equally-sized (but less
obtrusive) lower horn. This phenomenon agrees with similar findings in
a related study~\cite{degennaro}.

\begin{figure}[htb]
\centering
\subfigure[Favorably skewed horn.]{\includegraphics[width=.45\textwidth]{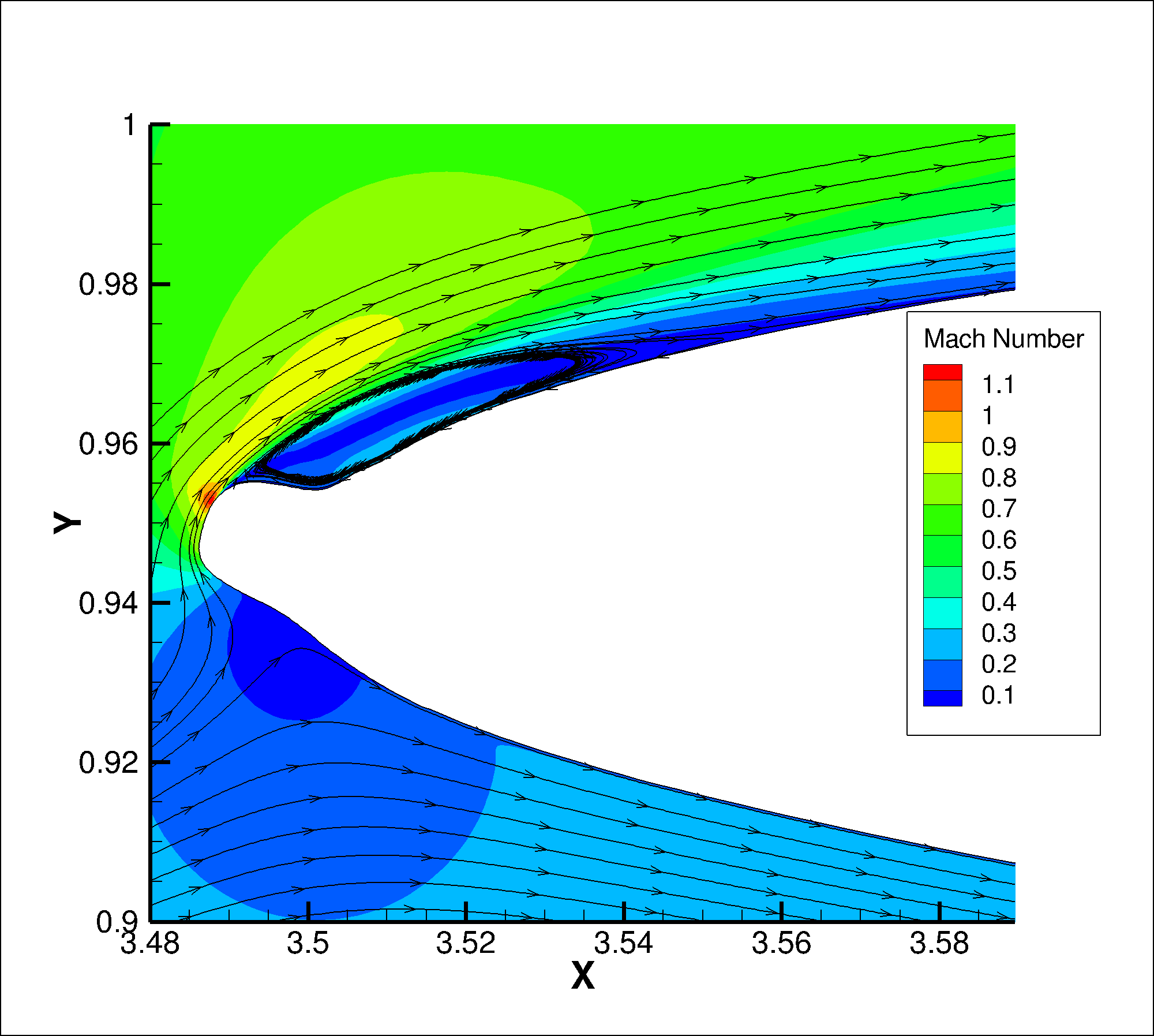}}
\subfigure[Unfavorably skewed horn.]{\includegraphics[width=.45\textwidth]{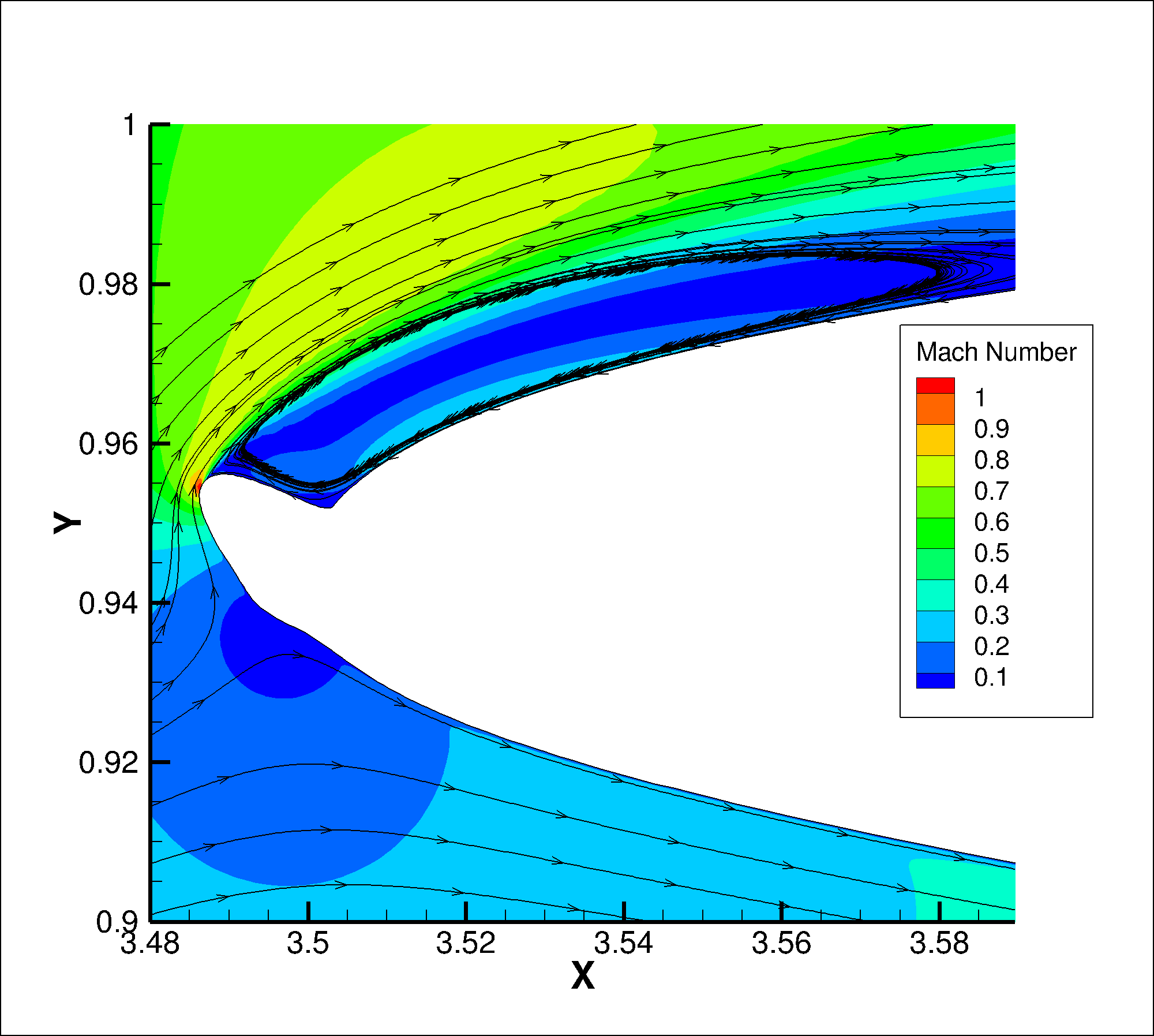}}
\subfigure[Contours of $\mathbb{E}(C_L | \text{POD1, POD2})$]{\includegraphics[width=.5\textwidth]{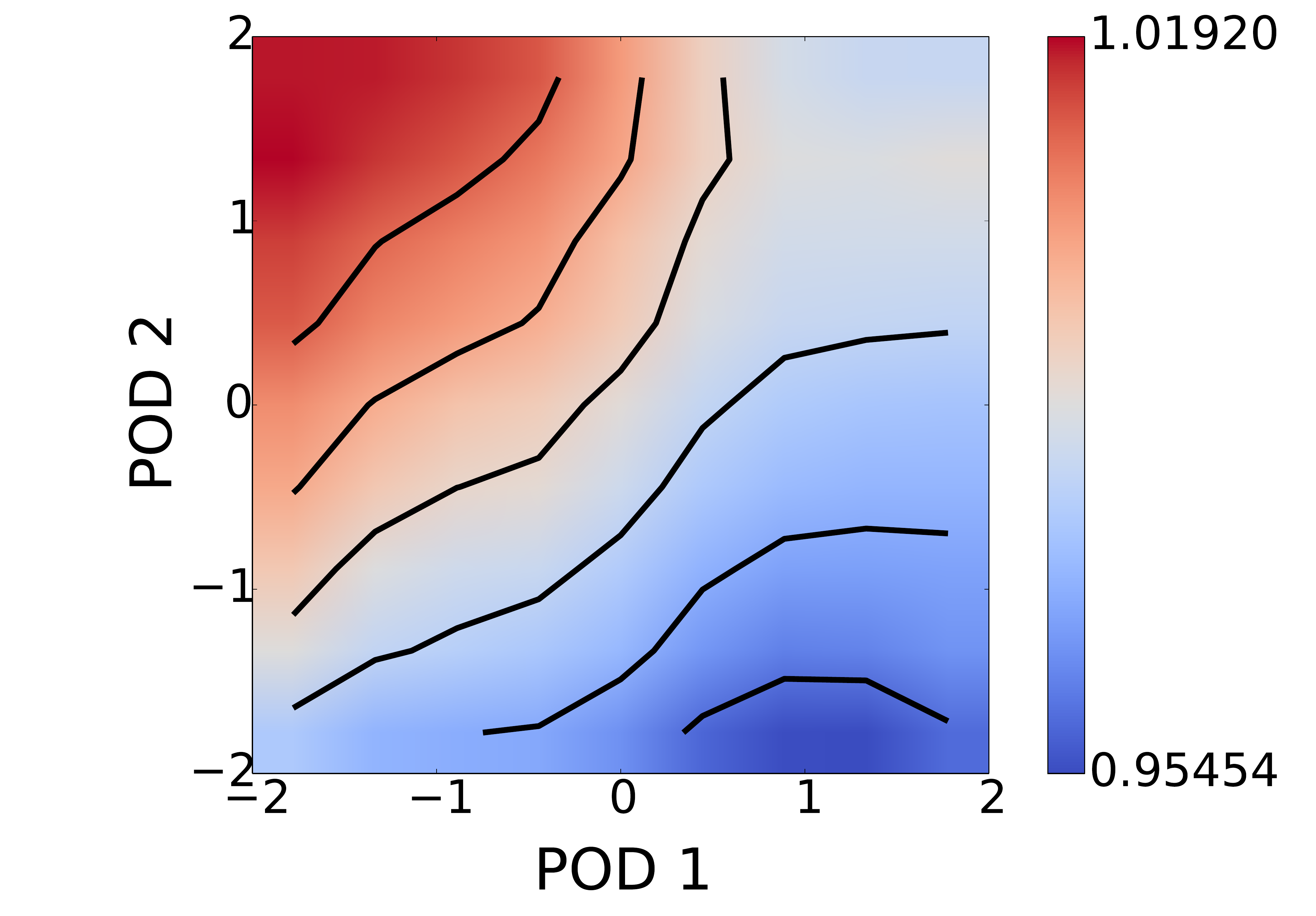}}
\caption{Effect of POD coefficients on performance. Part ($c$)
  displays contours of $C_L$ as a function of the POD coefficients,
  obtained by sampling the PCE surrogate and locally averaging over
  $b$ and $h$. Parts ($a$-$b$) display two horns that are the same
  size/position, but differ in their POD coefficients. This gives
  rise to favorable/unfavorable skewness in the shape.}
\label{fig:PODcontour}
\end{figure}

Although horn position is the most dominant parameter, the other
parameters do affect performance; this can be revealed by examining
the correlations in Table~\ref{tab:corr}. As expected, taller horns
give worse performance than shorter ones. Performance is relatively
insensitive to variations in width, but there is a slight tendency for
narrower horns to give worse performance than wider horns. This is
because narrower horns come to a sharper point and hence promote
larger leading edge separation bubbles, whereas wider horns have more
rounded points and hence are less severe. Perhaps not as immediately
clear is the effect of the POD modes on performance. One way to gain
some insight into this is to project the surrogate onto the
two-parameter space of POD coefficients. This can be approximated by
sampling the surrogate, and then locally averaging out the
three scaling parameters. This produces Fig.~\ref{fig:PODcontour},
which demonstrates that the POD modes can interact in such a way as to
produce a distinct skew to the horn. Depending on the direction of
this skew, this can either help or hurt aerodynamic performance, since
the length of the leading edge separation bubble depends on the horn
shape details.

\begin{figure}[htb]
\centering
\subfigure[Favorable horns.]{\includegraphics[width=.45\textwidth]{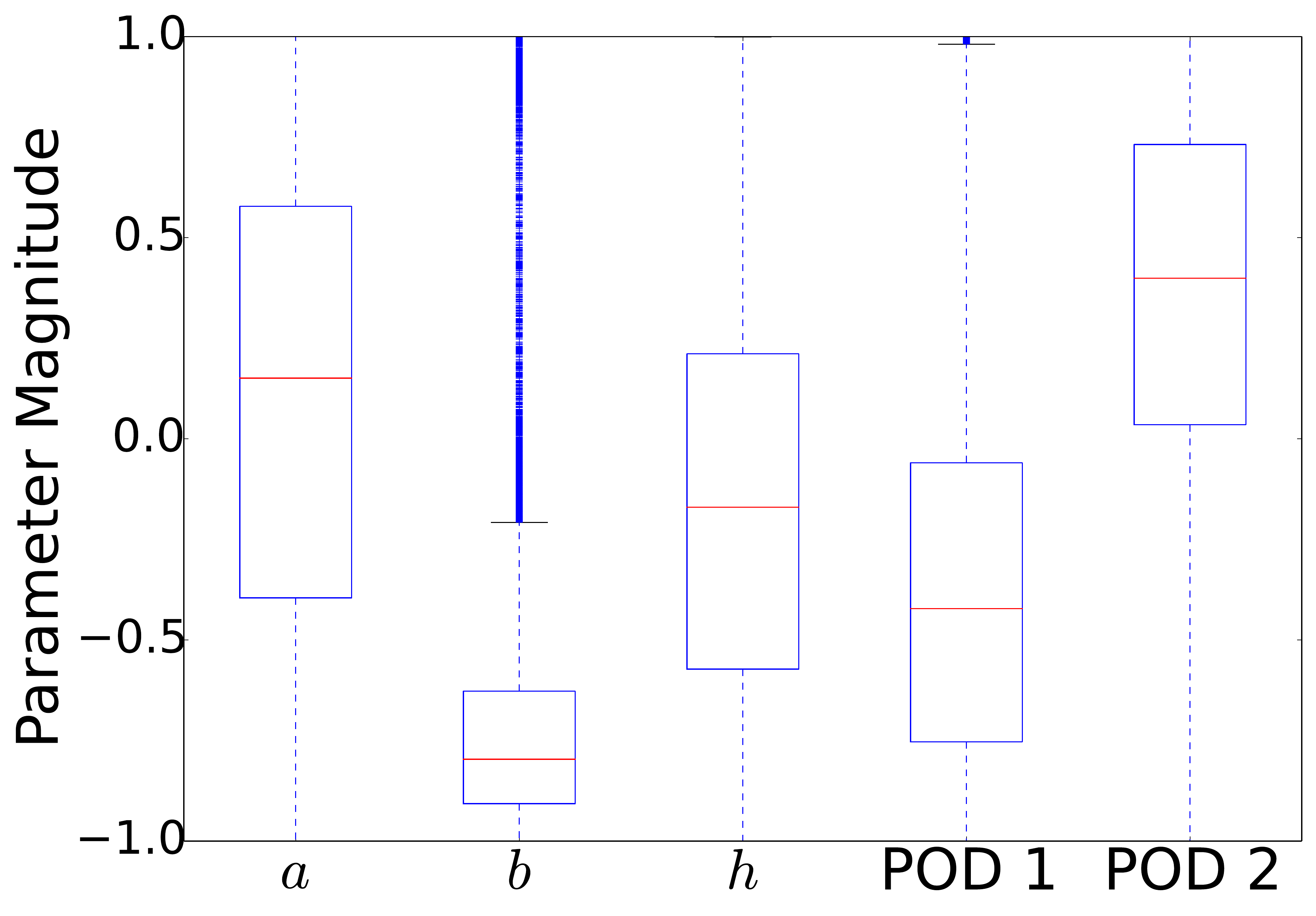}}
\subfigure[Unfavorable horns.]{\includegraphics[width=.45\textwidth]{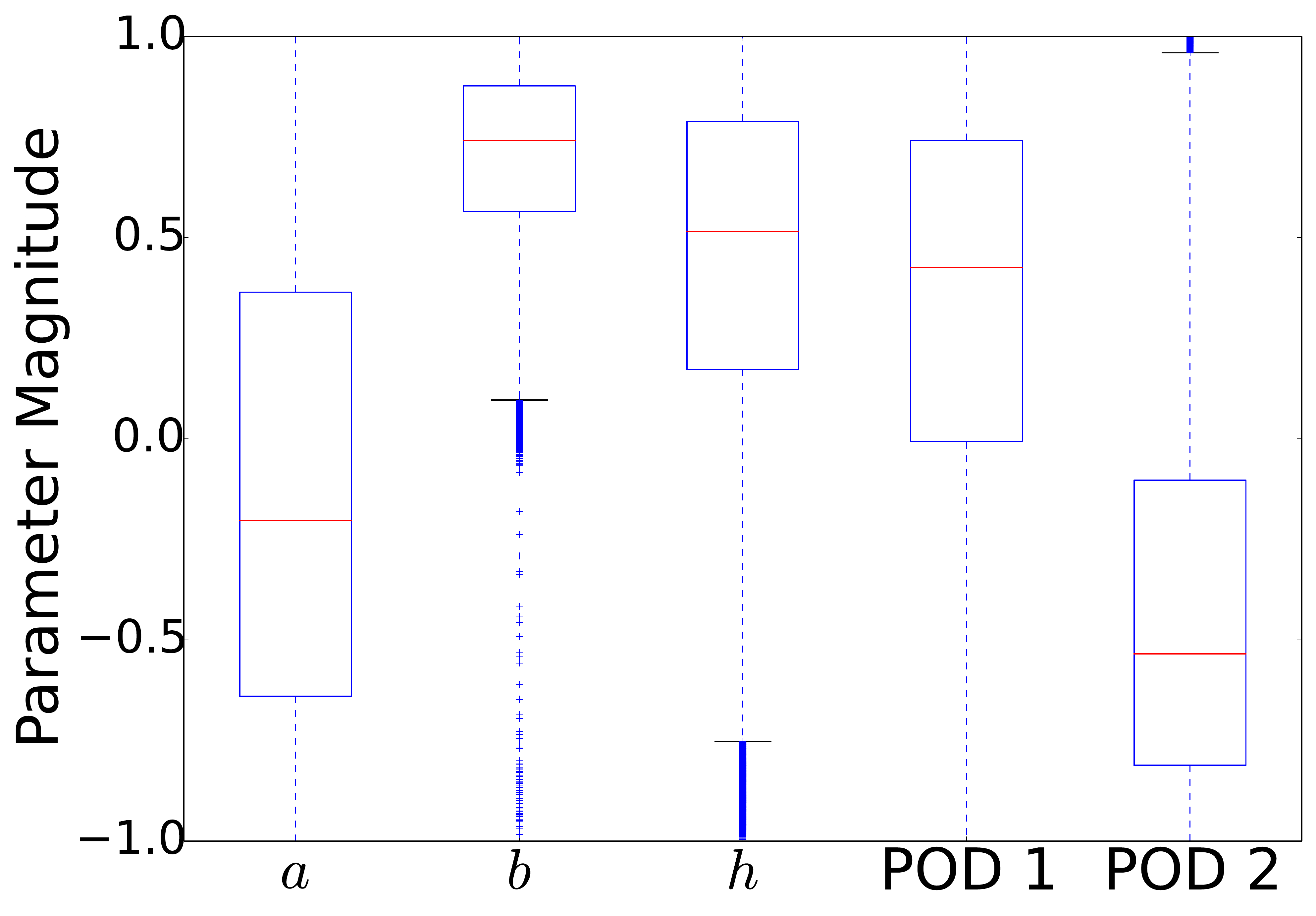}}
\caption{Statistical clustering of ``good'' (bottom 10$\%$ of $C_L$)
  and ``bad'' (top 10$\%$ of $C_L$) horns in parameter space, based on
  $10^6$ Latin Hypercube samples of the surrogate. The parameter
  magnitudes have all been linearly scaled to lie between $\pm 1$.}
\label{fig:GoodBadHornParamLocs}
\end{figure}

Integrating all of these analyses together gives a clean, intuitive
picture of the effects of our 5 parameters on the flow. We find that
the ``worst'' horn shapes in our parameter space are tall, narrow, upper
surface horns that have a sharp upper skew shape; the ``best'' ones
are short, wide and rounded, located on the lower surface, and have gentle downward
skew (or no skew at all). This is affirmed by examining
Fig~\ref{fig:GoodBadHornParamLocs}, which shows clear statistical
clustering of the horn shapes in parameter space that produce the best
and worst aerodynamic performance.

\sectionbreak{\clearpage}
\section{Results: Ice Shapes from Wind Tunnel Experiments}
\label{sec:results-expt}

Our goal in the preceding example was to use a dataset in which the
individual ice shapes exhibited only modest variation from the mean
shape. This ensured that we could obtain a faithful representation of
the ice using only a modest number of parameters (five). In this
section, we are interested in applying the same techniques to a
dataset whose entries represent a much wider range of physical
conditions, and hence a much wider range of shapes. As we will see,
doing this comes at the cost of having to retain more POD modes to
accurately represent the ice, which translates into a much larger UQ
study.

Two resources that provide an excellent sampling of 2D ice shapes for
different airfoil geometries subjected to different icing conditions
are Addy \cite{Addy:Shapes}, and Wright and Rutkowski
\cite{Wright:Shapes}. For the studies used in this example, we use a
subset of the shapes found in Addy. The shapes found in this database
were generated in a wind tunnel by exposing different airfoils to a
wide range of icing conditions. The conditions used reflect the
guidelines and standards for atmospheric icing conditions as defined
by the Federal Aviation Administration (Federal Aviation Regulations
25 Appendix C). In that database, three clean airfoil geometries are
used -- one which represents a business jet, one which represents a
commercial transport, and one which represents a general aviation
aircraft.

In this example, we limit the entries of our dataset to the business jet
ice shapes. We do this because having several different clean airfoil
geometries in our dataset would complicate our POD data reduction,
since in that case, one would not be able to uniformly define the
demarcations between the template (ie. the clean airfoil) and the
ice/air for all entries. Similarly, it would also make unclear the
choice of clean airfoil to use when generating artificial ice shapes
based on the POD modes. Consideration of the remaining two datasets in
Addy (the commercial transport and the general aviation aircraft)
would certainly make for interesting parallel studies, however, we
leave this as an item to be addressed in future work.

\begin{figure}[htb]
\centering
\includegraphics[width=0.75\textwidth]{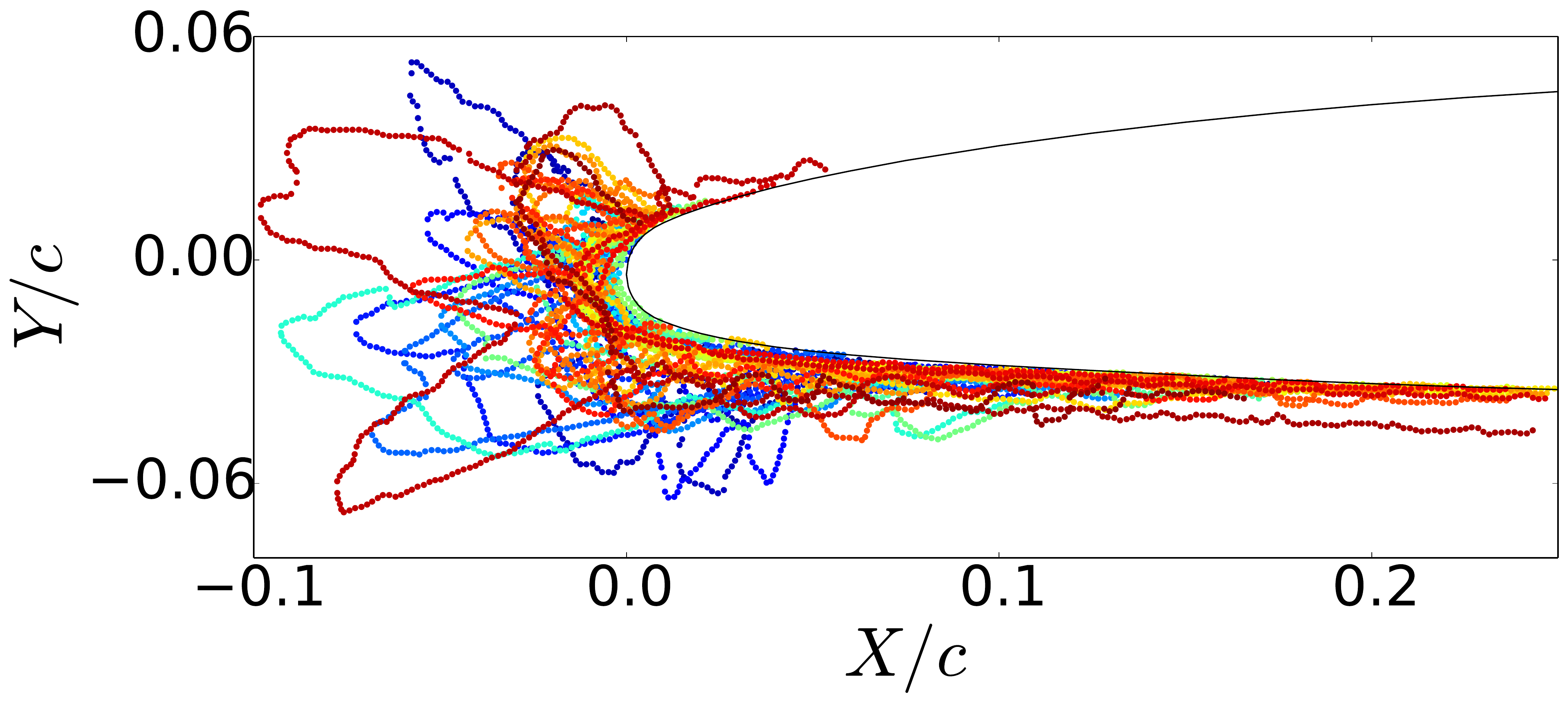}
\caption{Ice shape dataset, from Addy\cite{Addy:Shapes}. $c$ denotes
  the chord (set to 1 in all examples).}
\label{fig:Dataset}
\end{figure}

There were 54 total shapes from the business jet dataset. A plot of
all of these shapes together on the same airfoil is shown in
Fig. \ref{fig:Dataset}. As can be observed, there is significant
variation in the size and shape details of the ice. These shapes were
generated from variations in the following range of icing conditions:
\begin{itemize}
\item Mach number $\in [0.28,0.39]$
\item Airspeed $\in [175, 250]$ knots
\item Attitude $\in [1.5,6.0]^{\circ}$ 
\item Free-stream temperature $\in [-27.8,-0.7]^{\circ} C$ 
\item Surface temperature $\in [-31.6,-5.0]^{\circ}C$
\item MVD $\in [15,160] \mu m$ 
\item LWC $\in [0.310,1.6] \frac{g}{m^3}$
\item Exposure time $\in [0.7,45]$ min 
\end{itemize}
See Addy \cite{Addy:Shapes} (pg. 40) for further details.

\subsection{POD of the Ice Shape Data}

\begin{figure}[htb]
\centering
\includegraphics[width=0.75\textwidth]{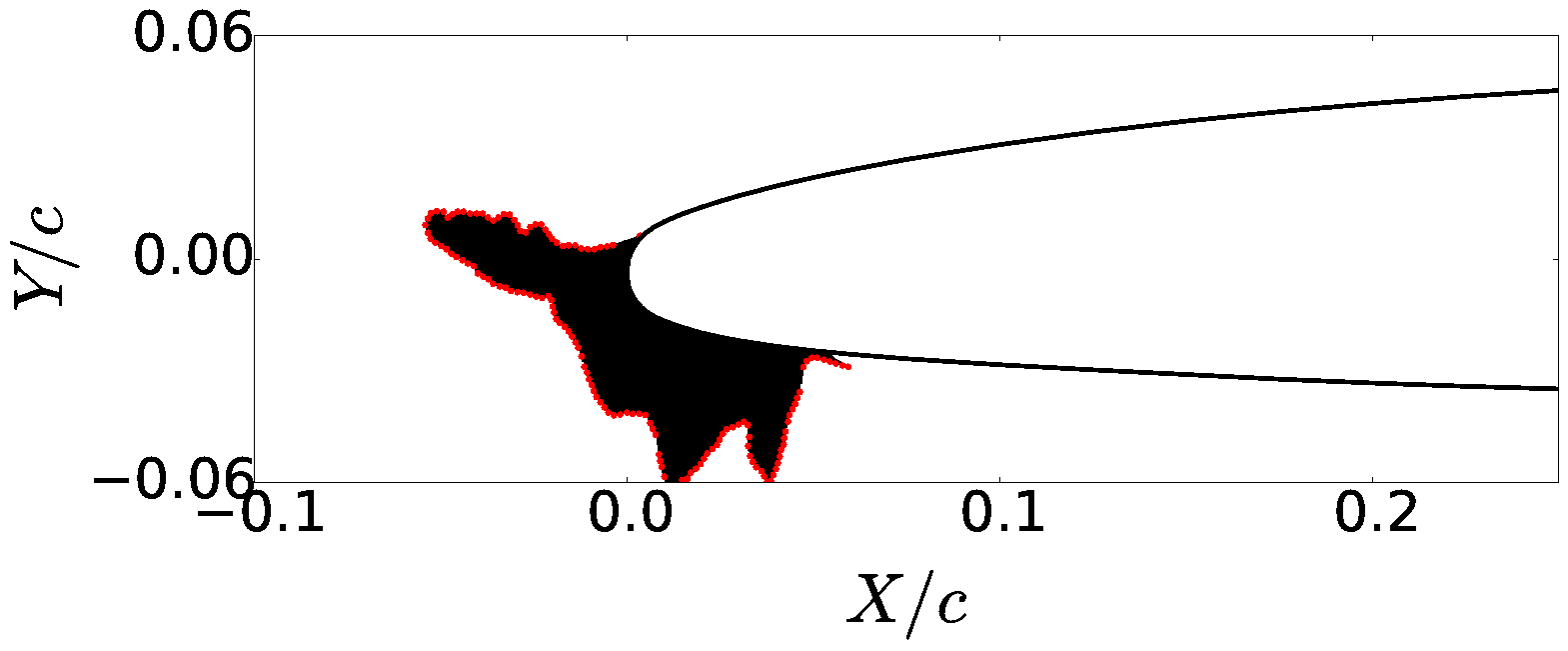}
\caption{Illustration of how ice shapes are defined for POD. Each ice
  shape is defined on a static Cartesian background mesh, the bounds
  of which form the rectangular window of this figure. For a
  particular ice shape, a grid point is assigned a value of 1 if that
  point is located inside the ice boundary. The ice boundary is shown
  in dotted red; points on the ice are shown in dotted black.}
\label{fig:DataVectorization}
\end{figure}

In order to apply the methods above to our ice shape problem, we need
to first determine an appropriate vector space representation of the
ice shapes. There is not necessarily one unique way of doing this. We
approach this problem by finding a rectangular window of space in
which all of the ice shapes in Fig. \ref{fig:Dataset} fit, and
overlaying this space with a static Cartesian mesh. For a particular
ice shape, we assign a value of 1 to a particular grid point if that
grid point is inside/on the body of the ice, or a value of 0 if it is
not. It should be noted that any points inside the clean airfoil were
excepted from the grid, so that our mesh consists entirely of points
located either in the free-stream, or in the ice. The background mesh
consists of roughly $N = 7\times10^5$ points. An example of this
process is shown in Fig. \ref{fig:DataVectorization}.

\begin{figure}[htb]
\centering
\includegraphics[width=0.6\textwidth]{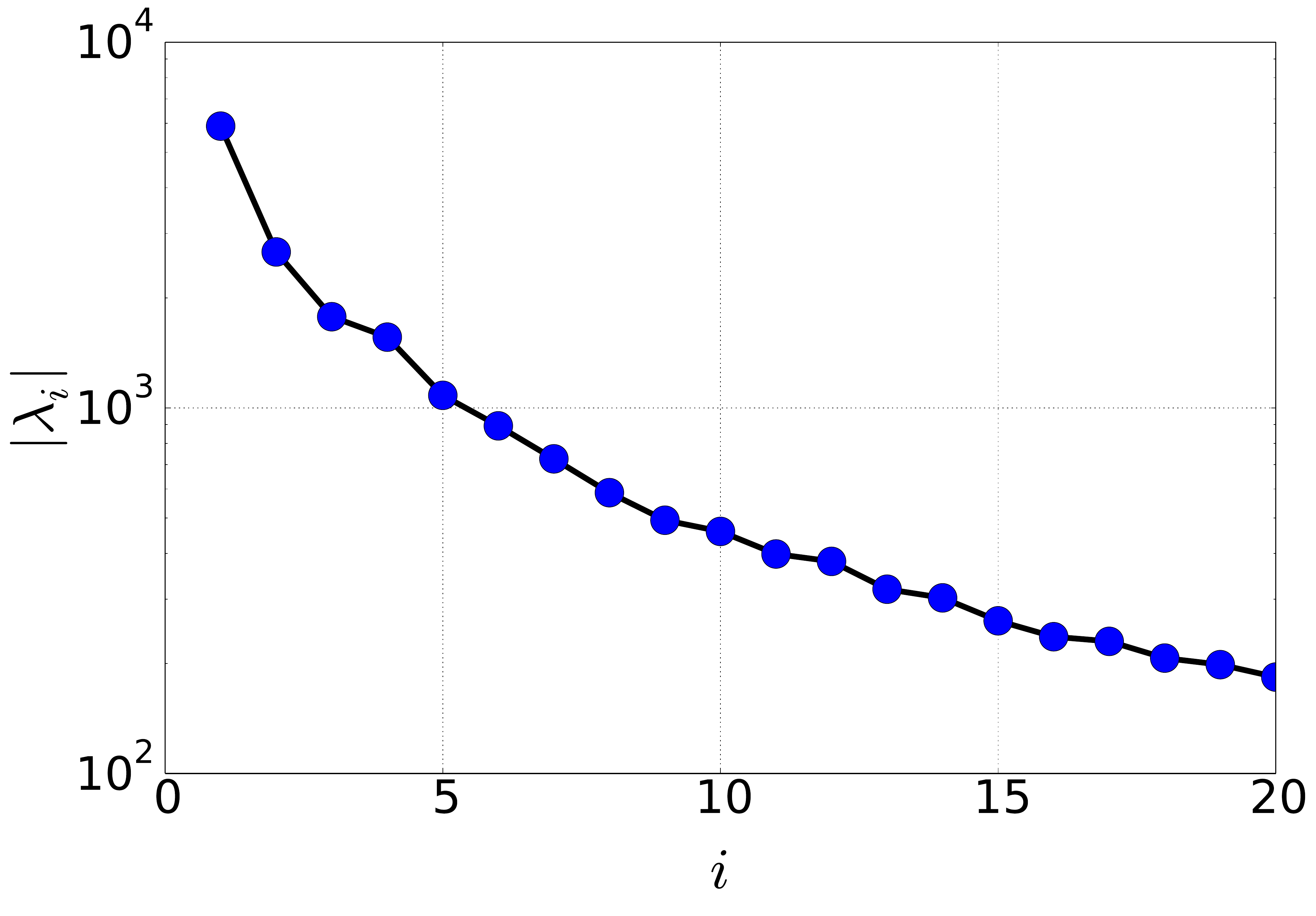}
\caption{Magnitudes of the POD eigenvalues.}
\label{fig:PODEvals}
\end{figure}

Having cast all of our ice shapes in the same $N$-dimensional vector
space, we are now in a position to compute the POD modes. Because we
will ultimately be doing UQ on the POD modes, we would like to retain
as few modes as possible while still having accurate representation
power in that basis. This is the classic tradeoff between economy and
accuracy in low-dimensional modeling. In order to make an informed
decision on how many modes to keep, we look at the magnitude of the
POD eigenvalues, shown in Fig. \ref{fig:PODEvals}.

We make the decision to truncate the expansion at order 8, where the
magnitudes of the eigenvalues have decayed by about one order of
magnitude. Retaining more modes would present a more computationally
laborious UQ study. Additionally -- as we will demonstrate shortly --
projection of the original dataset onto the first 8 modes yields good
reconstructions, so higher order modes may be practically unnecessary.

\begin{figure}[H]
 \begin{subfigmatrix}{2}
  \subfigure[Mean]{\includegraphics{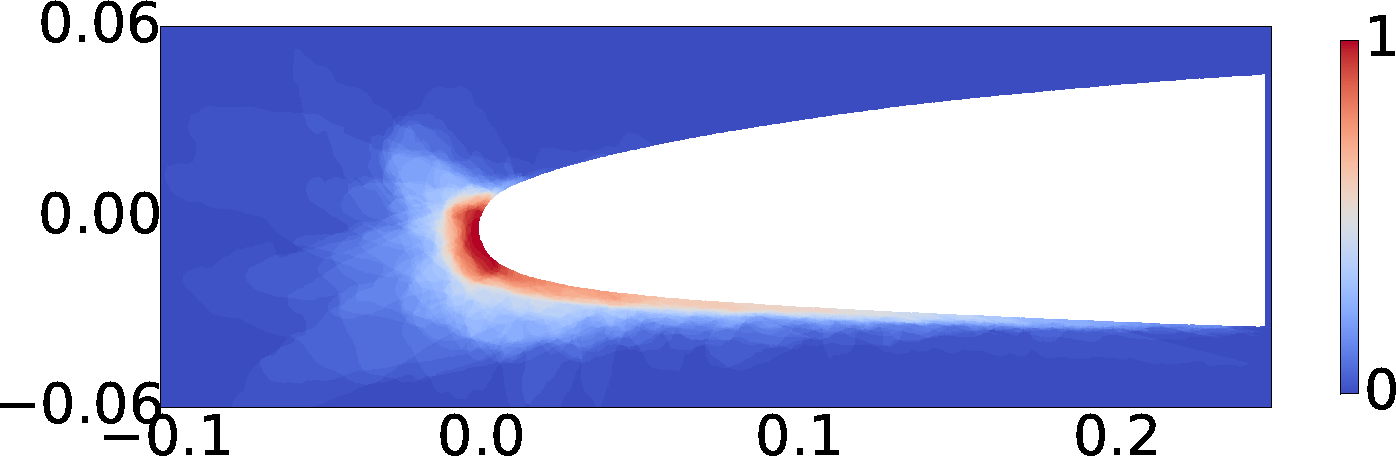}}
  \subfigure[Mode 1]{\includegraphics{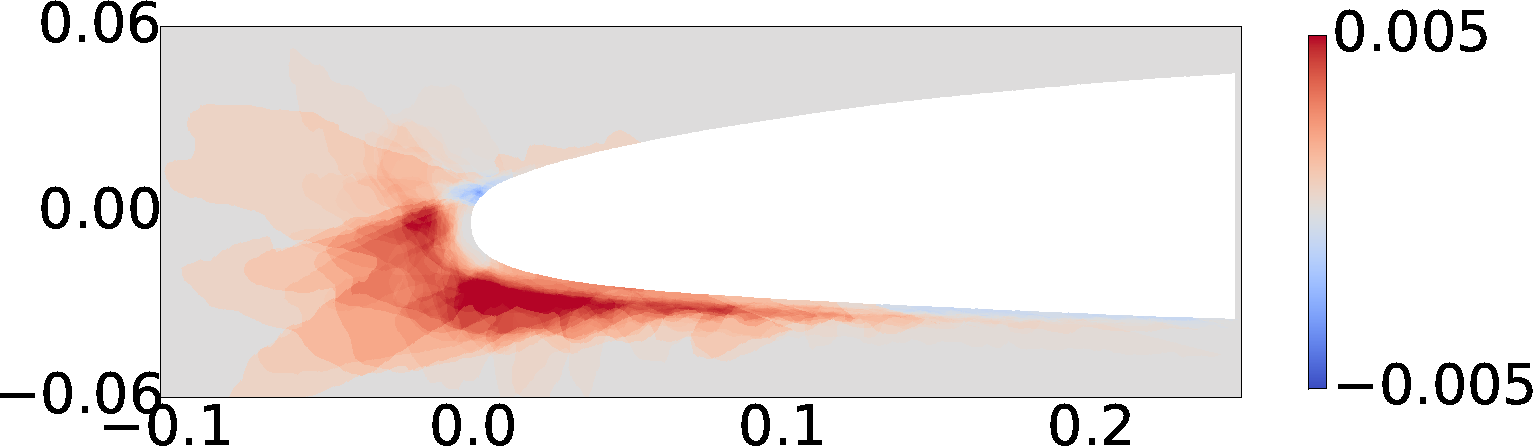}}
  \subfigure[Mode 2]{\includegraphics{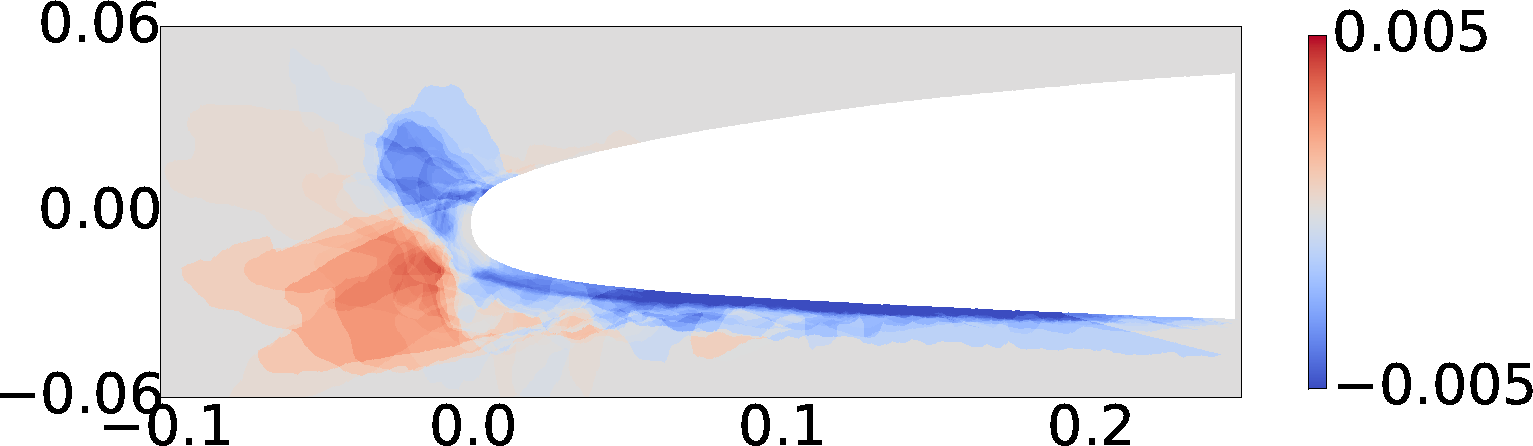}}
  \subfigure[Mode 3]{\includegraphics{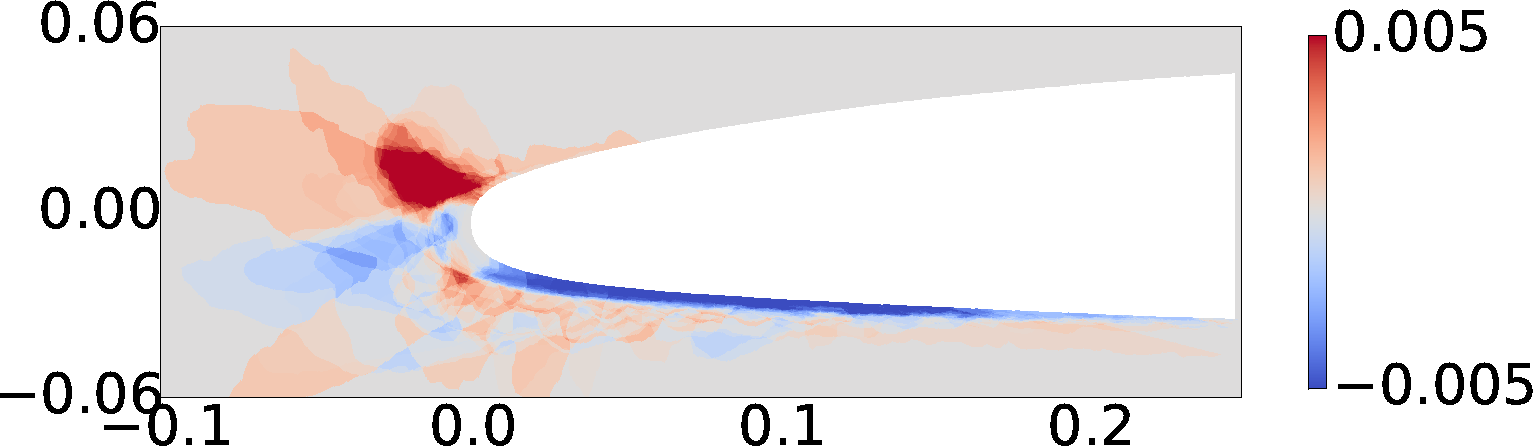}}
  \subfigure[Mode 4]{\includegraphics{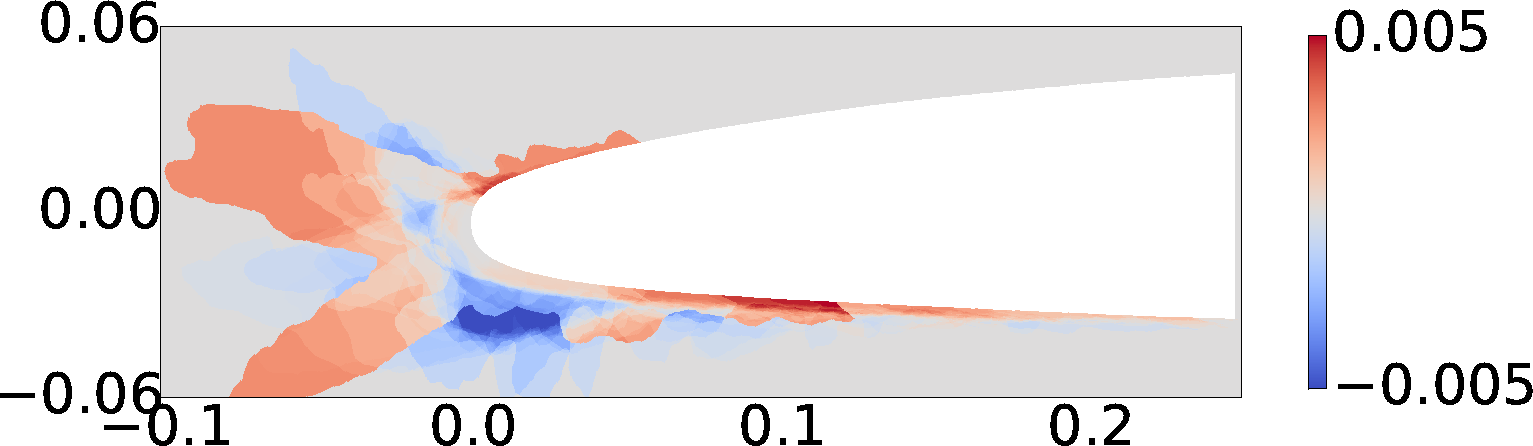}}
  \subfigure[Mode 5]{\includegraphics{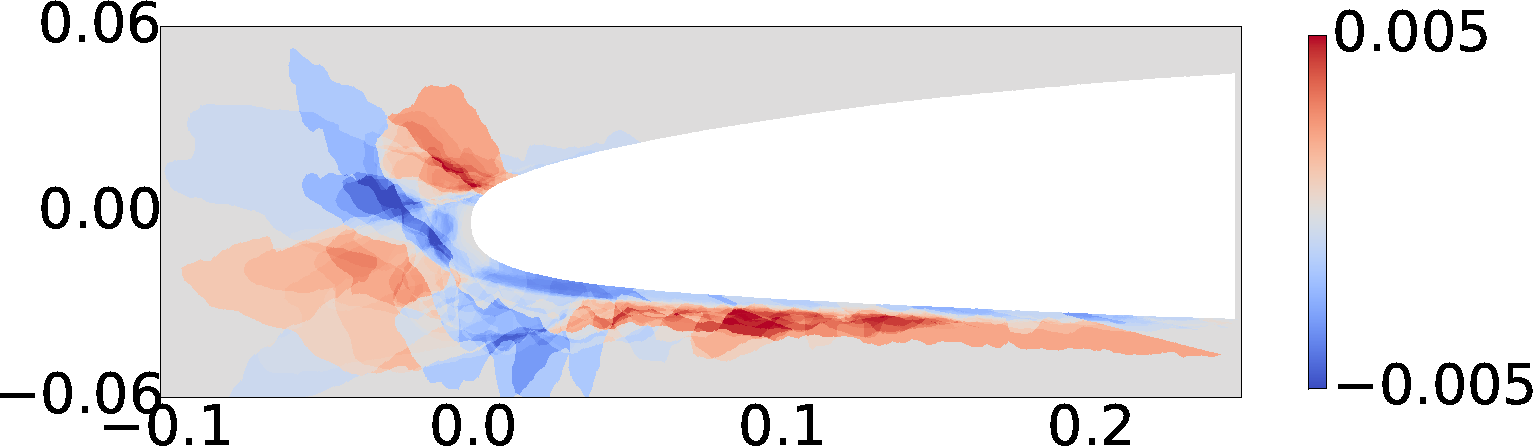}}
 \end{subfigmatrix}
 \caption{Mean and POD Modes.}
 \label{fig:PODModes}
\end{figure}

The mean and leading POD modes are shown in
Fig. \ref{fig:PODModes}. As can be seen, the mean and lowest order
modes have the most effect on the underside of the airfoil, where --
in our dataset -- there is a high probability of having ice (many of
our ice shapes have ``tails'' on the airfoil underside). The higher
order modes appear to be critical to attaining the extreme shapes in
our dataset. Intuitively, we might expect these modes to have a more
significant impact on the aerodynamics than the lower order ones.

\subsection{POD Reconstructions of the Ice Shapes}

We can now investigate how faithful our reconstructions of the
original dataset are using our POD basis. It is important to realize
that our POD reconstructions will need to be filtered {\it a
  posteriori}. This is because the reconstructions will, in general,
consist of grid point values that are between 0 and 1. However, our
ice shapes are binary in nature -- a particular grid point should
either be 1 if it is on the ice, or 0 if it is not.

\begin{figure}[H]
 \begin{subfigmatrix}{2}
  \subfigure[Unfiltered reconstruction]{\includegraphics{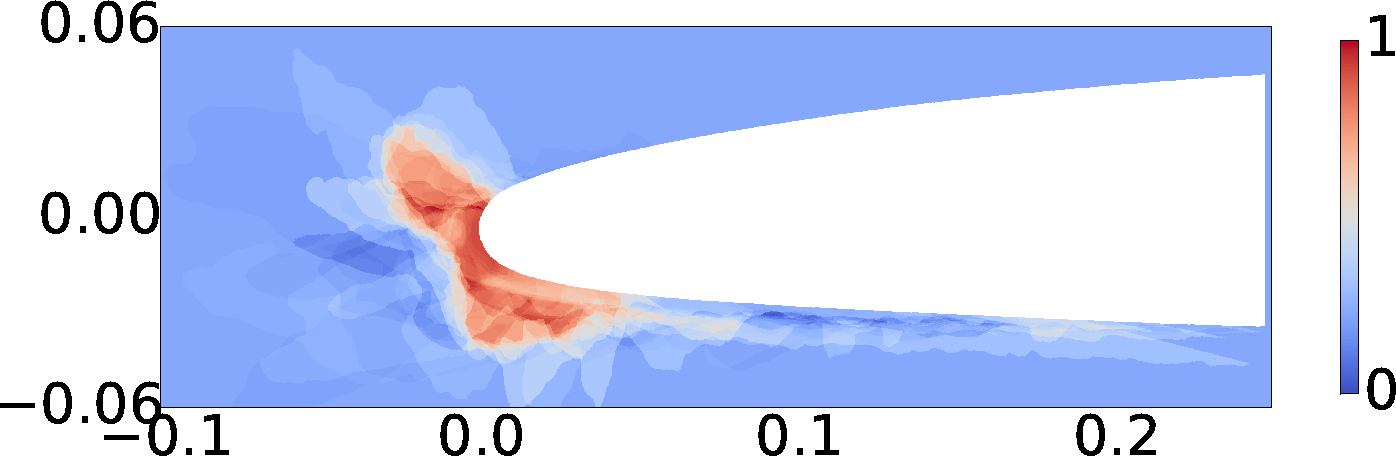}}
  \subfigure[Filtered reconstruction]{\includegraphics{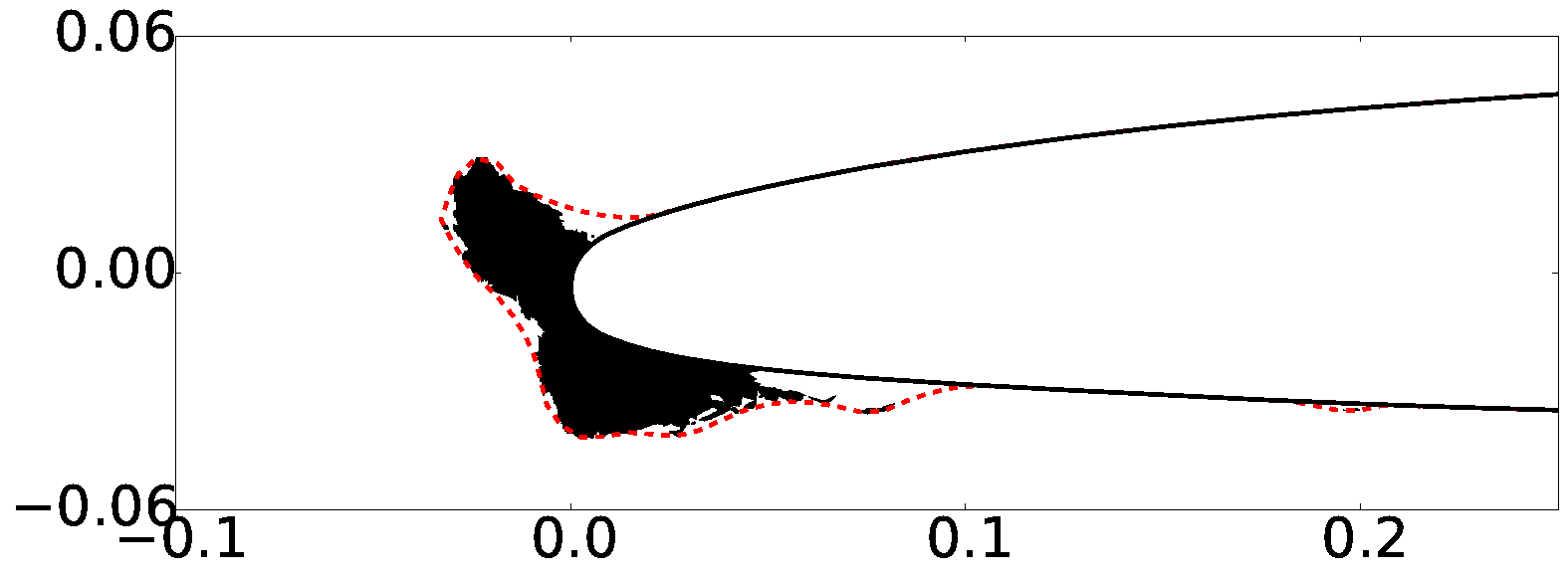}}
 \end{subfigmatrix}
 \caption{Unfiltered ({\it a}) and filtered ({\it b}) projections of
   an ice shape from the dataset onto the POD basis.}
 \label{fig:Filtering}
\end{figure}

\begin{figure}[H]
 \begin{subfigmatrix}{2}
  \subfigure[]{\includegraphics{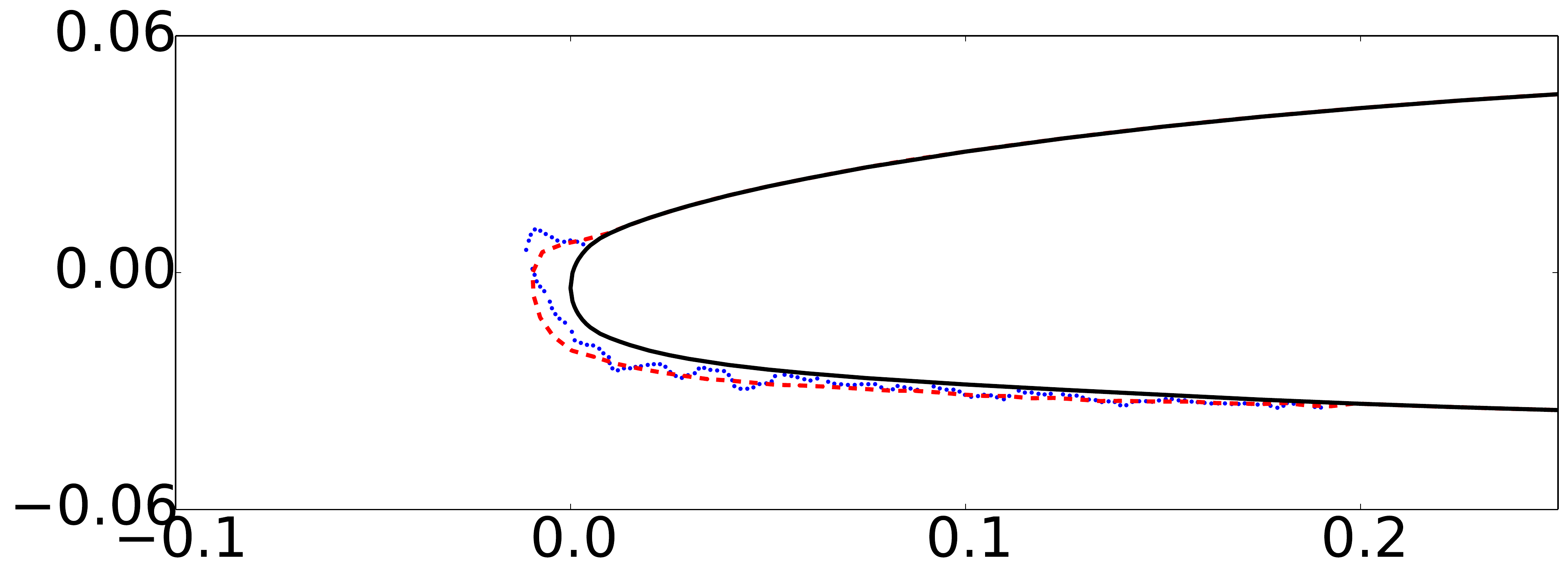}}
  \subfigure[]{\includegraphics{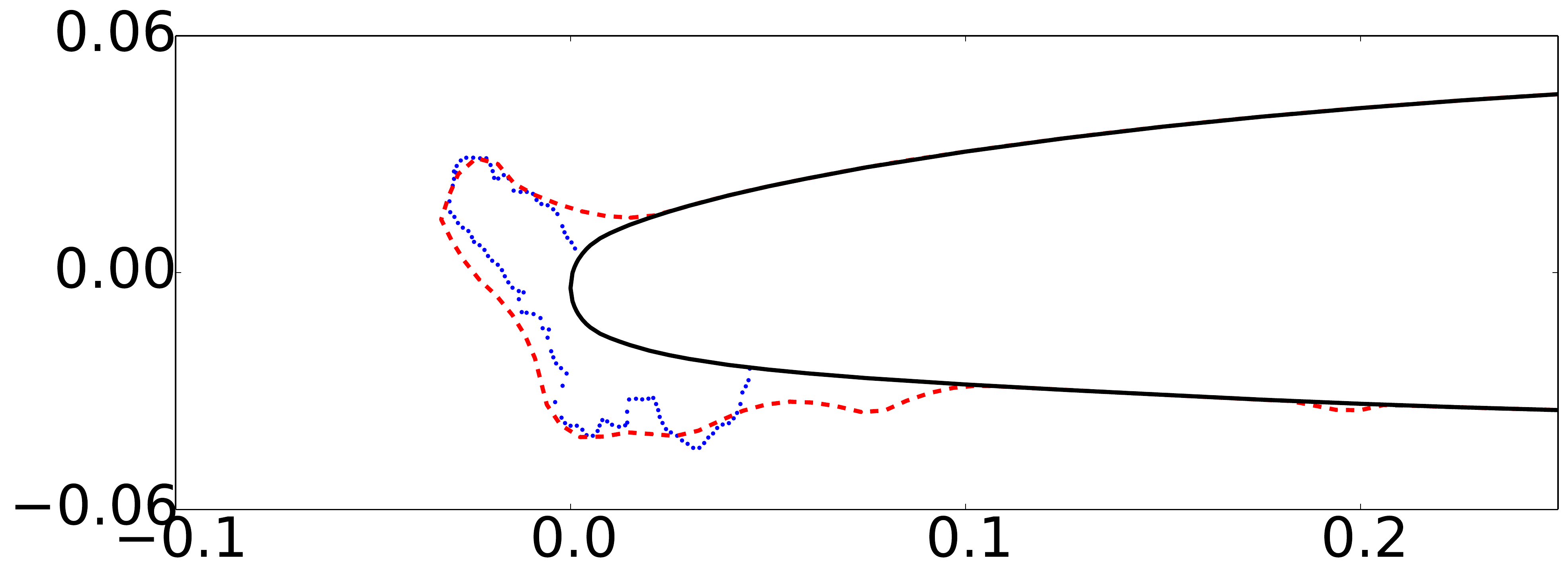}}
  \subfigure[]{\includegraphics{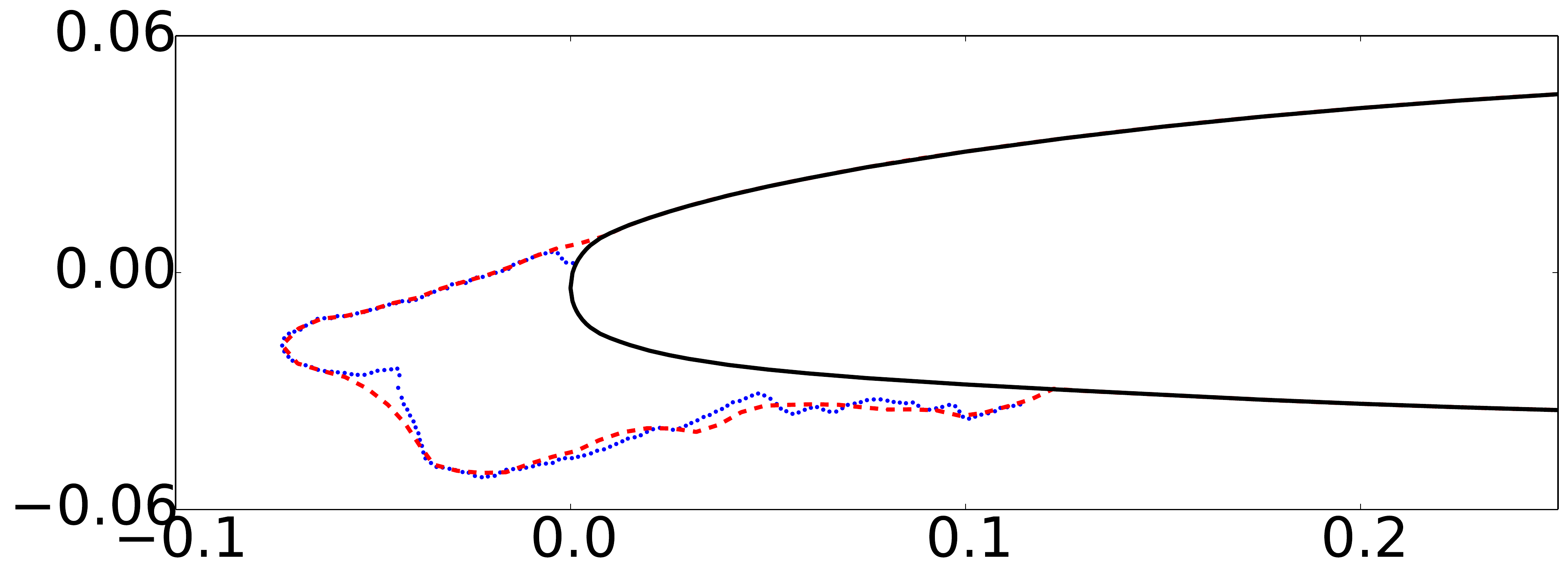}}
  \subfigure[]{\includegraphics{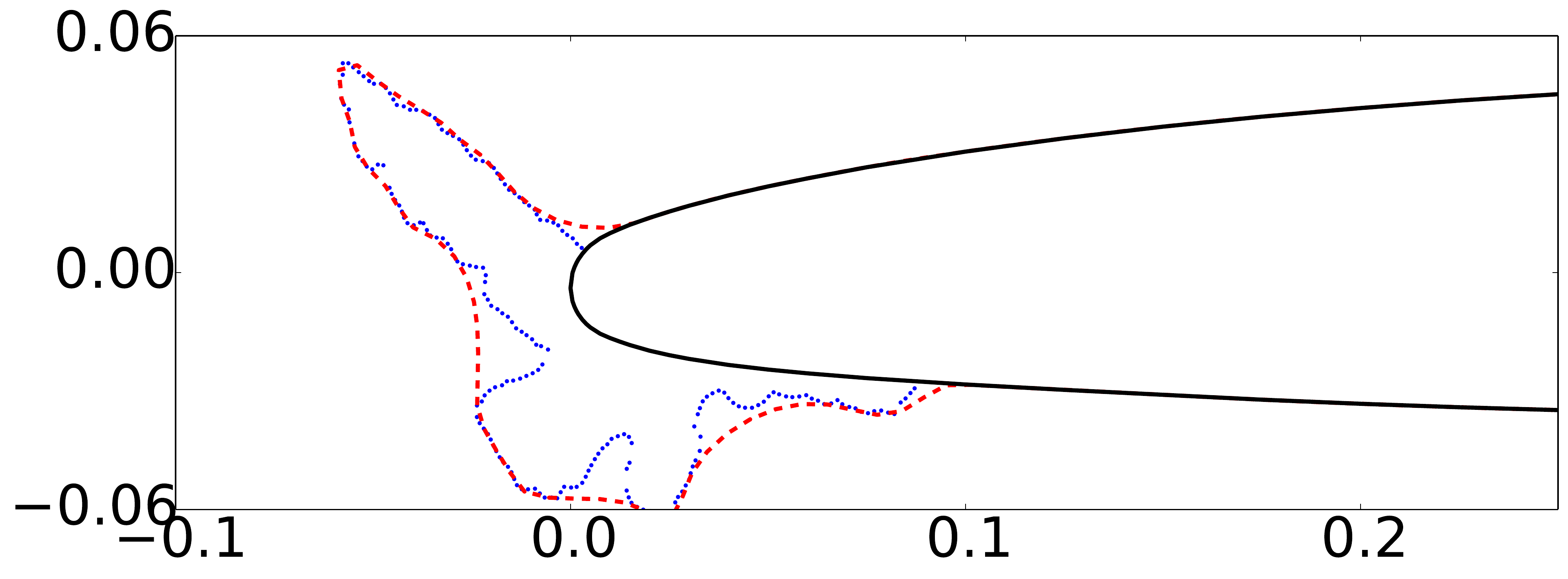}}
 \end{subfigmatrix}
 \caption{Original data ({\it blue}) compared to reconstructions ({\it
     red}) using 8 POD modes.}
 \label{fig:Reconstructions}
\end{figure}

To rectify this issue, we use a simple filter that rounds everything
less than 0.5 to 0, and everything greater than 0.5 to 1. An example
of this procedure is shown in Fig. \ref{fig:Filtering}. Several
filtered reconstructions are shown in
Fig. \ref{fig:Reconstructions}. As can be seen, the reconstructions
are of satisfactory agreement with the original data, which reaffirms
our choice of truncating the expansion at 8 modes.

\subsection{Airfoil Icing UQ: 8 Parameter Scenario}

The final step in this study is to apply the UQ techniques that we
have used in the previous example to this scenario, in which we have
an 8-dimensional parameter space that consists of variations of our
POD coefficients. Based on how many evaluations were required for the
5-parameter UQ study, we estimate that this UQ study may require
somewhere between five and ten-thousand flow solver evaluations to
yield a converged, faithful PCE surrogate. These simulations require two to four
weeks of wall-clock time, and are currently
underway.  We will report the results
when they become available.

\section{Conclusions}

Wing icing is not only dangerous to pilots, it is a complex physics problem
that is subject to a large amount of uncertainty. Quantifying the exact effects
of this uncertainty on airplane performance is hence of great importance to
airplane safety.

The purpose of this paper within that context is twofold: to
introduce a framework for low-dimensional modeling and data reduction
for ice shape datasets, and to introduce a fast, accurate, and
efficient means for quantifying uncertainty in the resulting shape
parameter spaces. Towards this end, we analyzed two examples, one in
which the dataset came from 2D spanwise cross-sections of a 3D wing
icing computation, and one in which the dataset came from icing wind
tunnel experiments performed at a wide range of icing conditions. In
the first example, we were able to effectively model the dataset using
only 5 parameters, successfully create a PCE surrogate, and analyze
the surrogate statistics for insights into the fundamental physics of
how the horn affects the aerodynamics. The second example is
demonstrative of how one can model more complicated ice shapes.
In that example, we were able to model the ice
shape dataset, but required 8 parameters to do so.

Much work and research remains to be done on these subjects in the
future. To begin, we need to complete our 8-parameter UQ study, which
will help enrich our understanding of the effects of more exotic ice
shapes on airfoil aerodynamics. Another possible avenue for research
is to first classify ice shapes into several distinct categories
(based on similarity in shape, similarity in the physical conditions
used to generate the shapes, etc.), and then to perform POD/UQ {\it
  separately} on each of these classes.
This approach would help alleviate the curse of dimensionality, by reducing the
number of parameters in each class.

\section*{Acknowledgments}
This work was supported by the FAA Joint University Program for Air
Transportation (JUP).
We would like to thank Brian Woodard at the University of Illinois at
Urbana-Champaign (UIUC) for sharing with us the LEWICE3D dataset used in this
paper.

\bibliographystyle{aiaa}	
\bibliography{GPCUQREF}		

\end{document}